\documentclass[draftcls,onecolumn,12pt]{IEEEtran}
\usepackage{bm}
\usepackage{epsfig}
\usepackage{stfloats}
\usepackage{graphicx}
\usepackage{subfigure}
\usepackage{epstopdf}
\usepackage{multirow}
\usepackage[sort, compress]{cite}
\usepackage{epsfig,graphics,subfigure}
\usepackage[cmex10]{amsmath}

\usepackage{array}
\usepackage{tabulary}
\usepackage{booktabs}
\usepackage{amsfonts}
\usepackage{algorithmic}
\usepackage{algorithm}
\usepackage{array}
\usepackage{tabulary}
\usepackage{booktabs}
\usepackage{amsfonts}

\usepackage{algorithmic}
\usepackage{algorithm}

\usepackage{epstopdf}
\def\BibTeX{{\rm B\kern-.05em{\sc i\kern-.025em b}\kern-.08em
    T\kern-.1667em\lower.7ex\hbox{E}\kern-.125emX}}

\usepackage{amssymb}

\usepackage{textcomp}
\usepackage{xcolor}

\begin{document}

\bibliographystyle{IEEEtran}

\title{Covert Transmission Assisted by Intelligent Reflecting Surface}
\author{\IEEEauthorblockN{Jiangbo Si, Zan Li, {\emph{Senior Member, IEEE}}, Yan Zhao,
Julian Cheng, {\emph{Senior Member, IEEE}}, Lei Guan, Jia Shi, and Naofal Al-Dhahir, \emph{Fellow, IEEE}}
\thanks{}
\thanks{Jiangbo Si,  Zan Li,  Yan Zhao, Lei Guan, and Jia Shi are with the Integrated Service Networks Lab
of Xidian University, Xi'an, 710071, China. Jiangbo Si is also a visiting scholar at the University of Texas at Dallas. (e-mail: jbsi@xidian.edu.cn).}
\thanks{Julian Cheng is with the School of Engineering, The University
of British Columbia, Kelowna, BC V1V 1V7, Canada (e-mail: julian.cheng@ubc.ca).}
\thanks{Naofal Al-Dhahir is with the Department of Electrical and Computer Engineering, The University of Texas at Dallas, Richardson,
TX 75080 USA.(e-mail: aldhahir@utdallas.edu).}
}

 \maketitle
\begin{abstract}
   Covert transmission is studied for an intelligent reflecting surface (IRS) aided communication system, where Alice aims to transmit messages to Bob without being detected by the warden Willie. Specifically, an IRS is used to increase the data rate at Bob under a covert constraint. For the considered model, when Alice is equipped with a single antenna, the transmission power at Alice and phase shifts at the IRS are jointly optimized to maximize the covert transmission rate with either instantaneous or partial channel state information (CSI) of Willie's link. In addition, when multiple antennas are deployed at Alice, we formulate a joint transmit beamforming and IRS phase shift optimization problem to maximize the covert transmission rate. One optimal algorithm and two low-complexity suboptimal algorithms are proposed to solve the problem. Furthermore, for the case of imperfect CSI of Willie's link, the optimization problem is reformulated by using the triangle and the Cauchy-Schwarz inequalities. The reformulated optimization problems are solved using an alterative algorithm, semidefinite relaxation (SDR) and Gaussian randomization techniques. Finally, simulations are performed to verify our analysis. The simulation results show that an IRS can degrade the covert transmission rate when  Willie is closer to the IRS than Bob.

\end{abstract}
\begin{IEEEkeywords}
Covert transmission, imperfect CSI, intelligent reflecting surface (IRS), joint beamforming and phase shift optimization.
\end{IEEEkeywords}
\IEEEpeerreviewmaketitle
\section{Introduction}

Exploiting reflecting surface is an energy and spectrum efficient technique to enhance the transmission rate and reliability of wireless communications. Different from the existing beamforming, multi-input multi-output and relay techniques,  intelligent reflecting surface (IRS) reflects the received signals without dedicated energy and additional signal processing functions, such as encoding, decoding, and modulation. The reflection is software controlled and realized by adjusting the amplitudes and phase shifts of the IRS elements. By proactively adjusting the amplitudes and phase shifts, the desired signals can be strengthened significantly and the undesired signals can be suppressed or cancelled effectively. Due to these advantages, IRS can be deployed densely in future cellular networks to improve the transmission coverage and reliability.

Recently, there have been several works focusing on IRS-aided transmission. In \cite{EBasar,Qwu3}, with or without channel knowledge, IRS was employed to improve single-input single-output (SISO) system performance. In a multi-input single-output (MISO) system, transmit beamforming, maximal ratio transmission (MRT), and IRS phase shifts were adapted according to only the channel state information (CSI) of the line of sight (LoS) components. The authors of \cite{MAlouini} extended the MISO system to the multi-input multi-output (MIMO) system, where the IRS reflecting elements and the precoder at the transmitter were alternatively optimized to minimize the symbol error rate. Moreover, the ergodic capacity and optimal phase shifts were computed based on the statistical CSI in \cite{SJin,DLi}. These research results show that IRS can boost the received signal-to-noise ratio (SNR) and enable ultra reliable communications at extremely low SNR.

In addition, IRS can be applied in multi-user systems. For example, in multi-user MISO downlink communications, transmit beamforming and IRS phase shifts were jointly optimized according to the instantaneous CSI of all links in \cite{CHuang}, where zero-forcing beamforming was applied at the transmitter to eliminate the multiuser interference and  the majorization-minimization algorithm was proposed to maximize the total transmission rate. For the same scenario, the authors in \cite{MAlouini2} studied the reflect beamforming design to maximize the minimum signal-to-interference-plus-noise ratio (SINR) with only the channel's statistical CSI. For the case of broadcasting messages to multiple users, the transmission beamforming vector at the base station (BS) and the phase shifts at the IRS were jointly optimized to minimize the total transmission power under quality of service (QOS) constraints at each user \cite{Hhan}. In addition, joint active beamforming at the transmitter and passive beamforming at the IRS was optimized by an iterative method in \cite{Qwu1}, where beamforming was designed  by applying the well-known minimum mean squared error (MMSE) criterion to cope with the multiuser interference. Subsequently, since the continuous phase shift is costly to be realized in practice, joint continuous transmit beamforming at the AP and discrete phase shifts reflect beamforming at the IRS was addressed in \cite{Qwu2}. It was shown that discrete IRS phase shifts incurred a constant performance loss that depends only on the number of phase shift levels.  Moreover, the IRS was used to assist the downlink transmission to cell-edge users by mitigating the inter-cell interference \cite{Hanzo}. All of the above studies show that the IRS can improve the transmission rate and decrease the energy consumption in multiuser systems.

Furthermore, IRS was also deployed to enhance the secrecy transmission performance. For an IRS-aided multi-input single-output single-antenna eavesdropper (MISOSE) system, the transmit beamforming vector at Alice and phase shifts at the IRS were jointly optimized for secrecy rate maximization in \cite{Gzhang} or transmission power minimization in \cite{JShi}, where an alterative optimization algorithm and semidefinite program (SDP) were used. The block coordinate descent (BCD) and minorization maximization (MM) techniques were also applied to solve the same optimization problem \cite{RSchober}. When multiple eavesdroppers and multiple legitimate users coexist in a MISO downlink communication system, joint active beamforming at the BS and passive beamforming at the IRS were optimized to maximize the minimum secrecy rate \cite{YLiang}. The authors of \cite{wxu,HWang} extended the optimization problem to multi-input multi-output multi-antenna eavesdropper (MIMOME) systems. In addition, a beamforming and artificial noise (AN) scheme was proposed in an IRS-aided system to enhance secrecy performance \cite{XGuan}.

Covert transmission is an emerging and cutting-edge communication security technique, which
guarantees a negligible detection probability at a warden \cite{7355562}. Covert transmission can protect confidential information from detection and can be used in systems with high security requirements, such as finance, national security, and military. There are many papers focusing on covert transmission \cite{7084182,8379465,7964713,Xliao}. However, few works have investigated IRS-aided covert transmission.  In \cite{HJiang}, the authors introduced the IRS technique in covert transmission, and outlined the possible applications of IRS in covert transmission systems, but they did not investigate IRS-aided covert transmission techniques or quantify their performance. For an IRS-aided covert transmission, on the one hand, the covert signal received at the legitimate receiver should  be strengthened by IRS reflection. On the other hand, the covert signal received at the warden should be weakened by IRS reflection. Hence, it is challenging to achieve a trade-off between strengthening the legitimate receiver's link and weakening the warden's link. This paper fills this gap and provides insights on IRS design in covert transmission. In addition, considering that multiple antennas is an adopted technology in the fifth-generation wireless (5G) systems, we consider the IRS-aided covert transmission with either a single antenna or multiple antennas at the transmitter. Furthermore, the covert transmission rate of an IRS-aided system is investigated under different kinds of CSI: instantaneous CSI of Willie's link, partial CSI of Willie's link, and imperfect CSI of Willie's link. The main contributions of this paper are summarized as follows:

\begin{itemize}
\item[1.] {For the first time, the performance of IRS-aided covert transmission is comprehensively investigated.} IRS is used to strengthen the signal power received at Bob and to weaken the signal power received at Willie. When Alice is equipped with a single antenna, a joint transmission power and phase shifts optimization problem is formulated to maximize the covert transmission rate. According to partial CSI of Willie's link, the optimal transmission power satisfying the covert constraint is first obtained. Then, the triangle inequality is used to derive the optimal phase shifts at the IRS. When full instantaneous CSI of Willie's link is available, the coupled transmission power and phase shifts are alternatively optimized for covert rate maximization. {Analysis and simulation results show that IRS is not necessarily beneficial to covert transmission, even if the partial or instantaneous CSI of Willie's link are available at Alice.}

\item[2.] When Alice is equipped with multiple antennas, a joint transmit beamforming and phase shifts optimization problem is formulated to maximize the covert transmission rate. When only partial CSI of Willie's link  is available, optimal transmit beamforming, namely MRT, is applied at Alice. Then, SDP is used to calculate the optimal IRS phase shifts. When all the channels' instantaneous CSI are available, optimal transmit beamforming with given phase shifts and optimal phase shifts with  given transmit beamforming are performed iteratively until the accuracy of the covert transmission rate is satisfied. Moreover, to decrease the algorithm's complexity, two suboptimal algorithms are proposed to minimize the signal power received at Willie. Simulation results show that the covert transmission rate without IRS assistance can be greater than that with IRS assistance. Moreover, although increasing the number of IRS elements can improve the covert transmission performance, the amount of increase is affected significantly by the channel quality of the IRS-to-Bob links.

\item[3.] Considering that Willie is often unfriendly to Alice, we investigate the impact of imperfect CSI of Willie's link on IRS-aided covert transmission. By using the triangle and the Cauchy-Schwarz inequalities, {we reformulate the optimization problem under the cases of imperfect CSI of the Alice-to-Willie link, imperfect CSI of the Alice-to-IRS link, imperfect CSI of the IRS-to-Willie link, and imperfect CSI of both the Alice-to-Willie and the IRS-to-Willie links.} The alternative optimization algorithm, semidefinite
relaxation (SDR), and Gaussian randomization techniques are used to solve the optimization problem and compute the maximum achievable covert transmission rate. We find that imperfect CSI of the Alice-to-Willie link degrades the covert transmission rate significantly.
\end{itemize}

    The remainder of this paper is organized as follows.
Section II describes the system model.  A joint transmission power and phase shifts optimization problem is proposed and solved with either partial CSI or instantaneous CSI of Willie's link in Section III. In Section IV, when Alice is equipped with multiple antennas, the transmit beamforming and IRS phase shifts are jointly optimized under both power and covert constraints. An optimal algorithm and two suboptimal algorithms are proposed to solve the problem. In Section V, the optimization problem is reformulated and solved with imperfect CSI of Willie's link. Numerical results are presented and discussed in Section VI. Finally, we conclude the paper in Section VII.


\emph{Notations}- $\left(  \cdot  \right)^T$ and $\left(  \cdot  \right)^H$ denote the transpose and conjugate transpose, respectively; The operator $\left|  \cdot  \right|$ denotes the absolute value; $\left\| {\left.  \cdot  \right\|} \right.$ denotes the Frobenius norm; $f_\upsilon\left(  \cdot  \right)$ is the probability density function (PDF) of  random variable (RV) $\upsilon$; $F_\upsilon\left(  \cdot  \right)$ is the cumulative distribution
function (CDF) of RV $\upsilon$; $\mathbb{E}{(\cdot)}$ and $Var{(\cdot)}$ denote the expected value and the variance of a RV, respectively; $Re{(\cdot)}$ and $Im{(\cdot)}$ respectively denote the real part  and the imaginary part of a complex RV; $\exp(\sigma^2)$ denotes exponential distribution having mean $\sigma^2$; $ arg\left(\bf{x}\right)$ stands for the phase vector of $\bf{x}$. $Tr\left(\bf{X}\right)$ denotes the trace of matrix $\bf{X}$. ${\lambda _{\min }}\left( {\bf{X}} \right)$ and  ${\lambda _{\max }}\left( {\bf{X}} \right)$,  respectively, denote the minimum and maximum eigenvalues of matrix ${{{\bf{X}}}}$; ${{{\mathbb{C}}}^{m \times n}}$ denotes the ${m \times n}$ complex number domain; ${\bf{I}}_M$ denotes the $M \times M$ identity matrix; ${\bf{x}} \sim \mathcal{CN}\left( {{\bf{\Lambda }},{\bf{\Delta }}} \right)$  denotes the circular symmetric
complex Gaussian vector with mean vector $\bf{\Lambda }$ and covariance matrix ${\bf{\Delta }}$. In addition, the symbol notations are given in Table I.
\renewcommand\arraystretch{1.4}{
\begin{table}
\caption{Summary of Notations}
\centering
\scriptsize
\begin{tabular}{|c|c|c|}
\hline
Notation & Description \\
\hline
${\bf{h}}_{a,j}$ & Instantaneous CSI between Alice and node $j$,  $j \in \{{Bob(b), Willie(w), IRS(s)}\}$ \\
\hline
${\bf{\tilde {h}}}_{a,j}$ & Imperfect CSI between Alice and node $j$,  $j \in \{{Willie(w), IRS(s)}\}$ \\
\hline
${\Delta\bf{{h}}}_{a,j}$ & Estimation error of ${\bf{h}}_{a,j}$   \\
\hline
${\bf{g}}_{s,j}$ & Instantaneous CSI between IRS and node $j$, $j \in \{{Bob(b), Willie(w)}\}$ \\
\hline
${\bf{\tilde {g}}}_{s,j}$ & Imperfect CSI between IRS and node $j$. $j \in \{{Willie(w)}\}$\\
\hline
${\Delta\bf{{g}}}_{s,j}$ & Estimation error of ${\bf{g}}_{s,j}$ \\
\hline
${\zeta_{i,j}}$ & The norm-bound of channel estimation errors between node $i$ and node $j$ \\
\hline
${\bf{\Theta}}$ & The diagonal reflecting matrix at the IRS  \\
\hline
$\bf{v}$ & The IRS phase shifts vector \\
\hline
$\theta_i$ & The phase angle of the $i$-th IRS element \\
\hline
$\bf{w}$ & Transmit beamforming vector at Alice \\
\hline
$\sigma^2_{j}$ & The noise covariance at node $j$,  $j \in \{{Bob(b), Willie(w)}\}$ \\
\hline
$\rho$ & The noise uncertainty coefficient at Willie  \\
\hline
$\lambda$ & The detection threshold at Willie \\
\hline
$\Gamma$ & The required accuracy for the covert transmission rate \\
\hline
$P_{D}$ & The transmission power at Alice without IRS assistance \\
\hline
$P_{I}$ & The transmission power at Alice with IRS assistance \\
\hline
${P_{\max }}$ & The transmission power constraint at Alice \\
\hline
$\xi$ &  The conditional detection error probability at Willie \\
\hline
$\kappa$ & A required threshold of detection error probability for covert transmission \\
\hline
$\eta$ & The maximum allowed received power at Wille to satisfy covert constraint \\
\hline
\end{tabular}
\end{table}}

\section{System Model}
\begin{figure}[ht]
\centering
\includegraphics[width=3.2in]{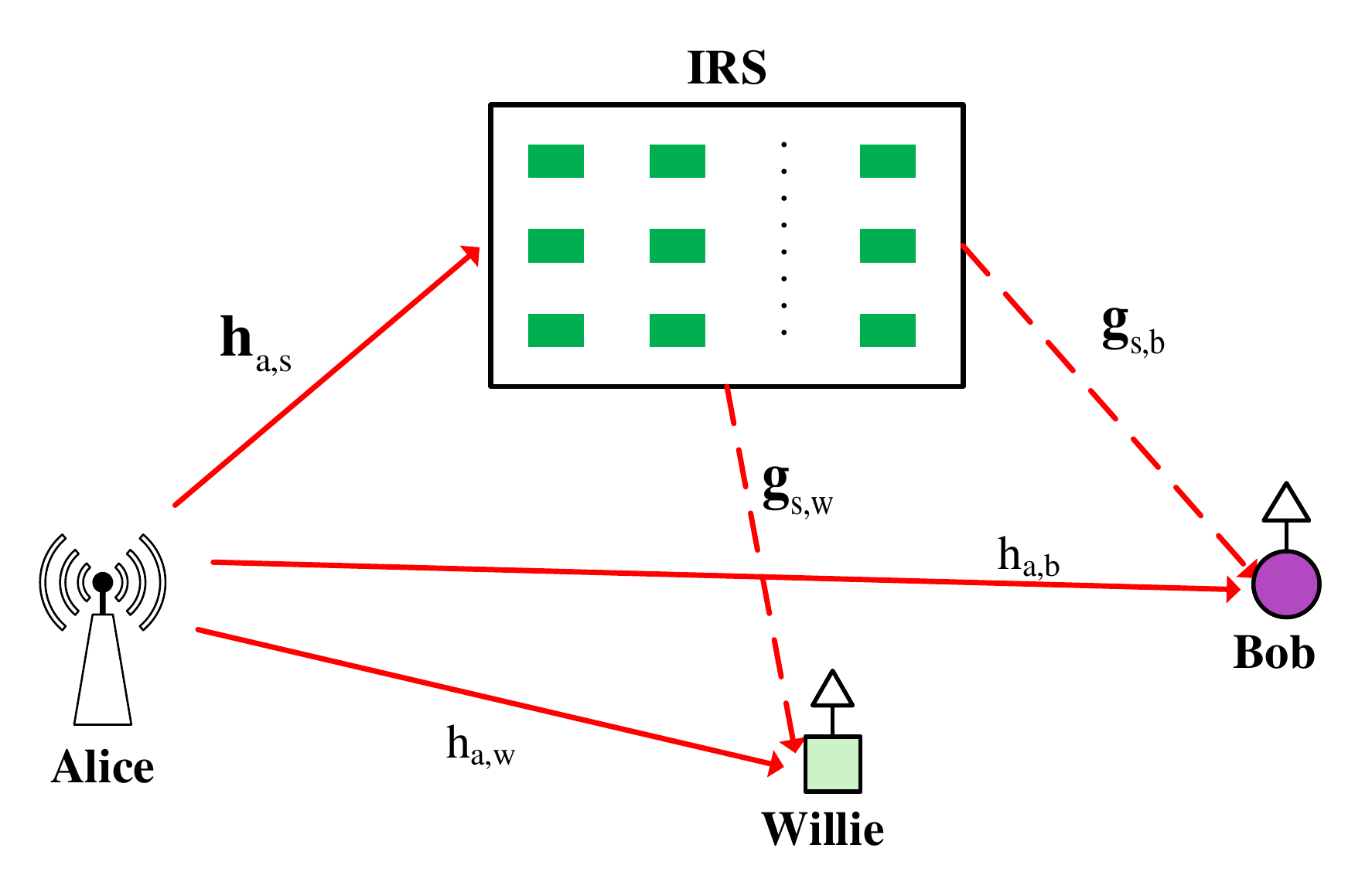}
\caption{Covert transmission with IRS assistance} \label{fig.1}
\end{figure}

As shown in Fig. 1, Alice wishes to transmit messages to Bob without being detected by Willie, who  is a warden  and tries to detect whether there exists a transmission from Alice or not. Moreover, Alice resorts to IRS with $N$ reflecting elements for improving the covert transmission rate. Both Bob and Willie are equipped with a single antenna, while $M$ antennas are deployed at Alice. The instantaneous CSI of the Alice $\rightarrow$ Bob and Alice $\rightarrow$ Willie links are, respectively, denoted by ${\bf{h}}_{a,b}\in \mathbb{C}^{M\times 1} \sim \mathcal {CN} \left( {{\bf{0}},\sigma^2_{a,b}{{\bf{I}}_{N}}} \right)$ and ${\bf{h}}_{a,w}\in \mathbb{C}^{M\times 1} \sim \mathcal {CN} \left( {{\bf{0}},\sigma^2_{a,w}{{\bf{I}}_{N}}} \right)$. In addition, ${{\bf{h}}_{a,s}}\in \mathbb{C}^{N\times M} \sim \mathcal {CN} \left( {{\bf{0}},\sigma^2_{a,s}{{\bf{I}}_{N}}} \right)$, ${\bf{g}}_{s,b}\in \mathbb{C}^{N\times 1} \sim \mathcal {CN} \left( {{\bf{0}},\sigma^2_{s,b}{{\bf{I}}_{N}}} \right)$, and ${\bf{g}}_{s,w}\in \mathbb{C}^{N\times 1} \sim \mathcal {CN} \left( {{\bf{0}},{\sigma^2_{s,w}}{{\bf{I}}_{N}}} \right)$ denote the instantaneous CSI of the Alice $\rightarrow$ IRS, IRS $\rightarrow$ Bob, and IRS $\rightarrow$ Willie links, respectively. The diagonal reflection matrix for the IRS is ${\bf{\Theta}}  = diag\left( {{e^{j{\theta _1}}},{e^{j{\theta _2}}}, \cdots ,{e^{j{\theta _N}}}} \right)$, where the amplitude of each element is assumed to be equal to $1$.  The maximum transmission power at Alice is $P_{max}$.

For covert transmission, we assume that Willie employs an energy detection method to determine whether there exists a transmission from Alice. Due to the electromagnetic environment variations, there is noise uncertainty at Willie, which affects the covert rate significantly. When the bounded uncertainty model is applied, the PDF of the noise power at Willie is given by \cite{JS2,BHe}
\begin{align}\label{distribution}\
{f_{{\sigma ^2}_w}}\left( x \right) = \left\{ \begin{array}{l}
\frac{1}{{2\ln \left( \rho  \right)x}},\frac{1}{\rho }{{\tilde \sigma }^2}_w \le x \le \rho {{\tilde \sigma }^2}_w\\
\begin{array}{*{20}{c}}
{0,}&{}&{\begin{array}{*{20}{c}}
{}&{otherwise}
\end{array}}
\end{array}
\end{array} \right.
\end{align}
where $\rho$ is the noise uncertainty coefficient,  and  ${\tilde \sigma ^2}_w$ denotes the noise power without noise uncertainty, { i.e. $\rho=1$}. The noise power at Willie satisfies the following inequality
\begin{align}\label{noiseuncertain}\
\frac{1}{\rho }{\tilde \sigma ^2}_w \le {\sigma ^2}_w \le \rho {\tilde \sigma ^2}_w.
\end{align}
In addition, since Bob does not detect the signal, we assume that the noise power at Bob is the additive noise (AWGN) with zero mean and fixed variance $\sigma^2_{b}$. The covert transmission is evaluated under two events: $H_0$ and $H_1$, where $H_0$ denotes the null hypothesis that Alice does not
transmit messages to Bob, and $H_1$ denotes the alternative hypothesis that Alice
transmits messages to Bob. In this paper,  we assume that the transmission is covert when the conditional detection error probability (DEP), which is defined as the summation of the miss detection probability under $H_1$ and the false alarm probability under $H_0$, is larger than a predetermined threshold $1-\kappa$ \cite{BHe,HQ}.

\section{Single Antenna at Alice}
In this section, we investigate IRS-aided covert transmission with a single antenna deployed at Alice ($M=1$). The covert transmission rate for an IRS assisted system is studied when either the instantaneous or partial CSI of Willie's link is known to Alice. Under both cases, considering that Bob is always friendly to Alice, we assume that the full instantaneous CSI of  Bob's link is perfectly known at Alice.
\subsection{With Partial CSI of  Willie's Link}
Note that without a covert constraint, the transmission rate at Alice is given by
\begin{align}\label{rate}\
{R_a} = \log 2\left( {1 + \frac{{{P_I}\left| {{h^H}_{a,b} + {{\bf{g}}^H}_{s,b}{\bf{\Theta }}{{\bf{h}}_{a,s}}} \right|}}{{{\sigma ^2}_b}}} \right)
\end{align}
Neglecting the logarithmic function and constant variables of \eqref{rate}, the optimization problem to maximize the covert constraint is formulated as
 \begin{subequations}
\begin{align}\label{singleproblema}\
&\mathop {\max }\limits_{P,\bf{\Theta} } {P_I\left| {{h^H_{a,b}} + {{\bf{g}}^H_{s,b}}{\bf{\Theta}}{{\bf{h}}_{a,s}} } \right|^2} \\
&{s.t.}\quad {P_I \le {P_{\max }}}  \label{singleproblemb}\\
&\quad\quad {{P_{FA}}   + {P_{MD}}  \ge 1 - \kappa}\label{singleproblemc}
\end{align}
 \end{subequations}
where the transmission power at Alice $P_I$ and the diagonal phase shifts matrix $\bf{\Theta}$ are jointly optimized for covert rate maximization. Equation \eqref{singleproblema} denotes the optimization goal: maximizing the received signal power at Bob, \eqref{singleproblemb} denotes the total transmit power constraint at Alice, and the covert transmission constraint is given in \eqref{singleproblemc}, where ${P_{FA}}$ and ${P_{MD}}$ denote, respectively, the false alarm probability and the miss detection probability, and will be defined shortly. When only partial CSI of Willie's link is available, the IRS phase shifts are optimized according to instantaneous CSI of  Bob's link. Then, optimal phase shifts for Bob's link are random for Willie's link, and the covert constraint is only determined by the transmission power $P_I$.  Hence, to solve optimization problem \eqref{singleproblema}, we first derive the maximum permitted transmission power, denoted by $P^*_I$, which is determined by the covert constraint.

Covert transmission is often evaluated under two events: $H_0$ and $H_1$. For the two events, the average signal power received at Willie is given by
\begin{align}\label{tnh0h1}\
{T_w} = \left\{ {\begin{array}{*{20}{c}}
{\sigma _w^2}&{{H_0}}\\
{P_I{\left| {{h^H_{a,w}} + {{\bf{g}}^H_{s,w}}\bf{\Theta}}{{\bf{h}}_{a,s}} \right|^2} + \sigma _w^2}&{{H_1}}.
\end{array}} \right.
\end{align}
According to \eqref{tnh0h1}, the probability of missed detection ${P_{MD}}$ and the probability of false alarm  ${P_{FA}}$ are, respectively, given by
\begin{align}\label{falsealarm}\
{P_{FA}} = \Pr \left( {{T_w} \ge \lambda |{H_0}} \right) = \Pr \left( {{\sigma ^2}_w \ge \lambda|{H_0} } \right)
\end{align}
and
\begin{align}\label{missdetection}\
{P_{MD}} = \Pr \left( {\left. {{T_w} \le \lambda } \right|{H_1}} \right) = \Pr \left( {P_I{{\left| {{h^H_{a,w}} + {{\bf{g}}^H_{s,w}}\bf{\Theta}}{{\bf{h}}_{a,s}}\right|}^2} + {\sigma ^2}_w \le \lambda |{H_1}} \right)
\end{align}
where $\lambda$ is the detection threshold at Willie. { Then, under the assumption that the apriori probability of
 hypotheses either $H_0$ or $H_1$ being true is 0.5 \cite{7964713},  the conditional DEP is}
\begin{align}\label{gamma_M}\
\xi  &= {P_{MD}}   + {P_{FA}} \nonumber \\
     &=1 - \Pr \left( {\lambda  -P_I{{\left| {{h^H_{a,w}} + {{\bf{g}}^H_{s,w}}\bf{\Theta}}{{\bf{h}}_{a,s}} \right|}^2}  < {\sigma ^2}_w < \lambda } \right).
\end{align}
It is noted that when $\lambda  < \frac{1}{\rho }{\tilde \sigma ^2}_w$, { ${P_{FA}}=1$ and ${P_{MD}}=0$. By contrast, when $\lambda  > {\rho }{\tilde \sigma ^2}_w$, ${P_{FA}}=0$ and ${P_{MD}}$ increases with $\lambda$}. Thus, from Willie's perspective,  to minimize the DEP,  the optimal detection threshold $\lambda$ should satisfy the following constraint
\begin{align}\label{constraint}\
 \frac{1}{\rho }{\tilde \sigma ^2}_w \leq \lambda \leq {\rho }{\tilde \sigma ^2}_w.
 \end{align}
 Let $z={{\left| {{h_{a,w}} + {{\bf{h}}_{a,s}}{\bf{\Theta}}{{\bf{g}}_{s,w}}} \right|}^2}$,  then, the average DEP conditioned on $\lambda$ is written as
\begin{align}\label{adep}\
\bar \xi  \left( \lambda \right) = 1 - \int_{\max \left( {\lambda  - {{zP_I}},\frac{1}{\rho }{{\tilde \sigma }^2}_w} \right)}^{\min \left( {\lambda ,\rho {{\tilde \sigma }^2}_w} \right)} {{f_{{{\sigma ^2_w}}}}\left( x \right)} dx = 1 - \int_{\max \left( {\lambda  - {{zP_I}},\frac{1}{\rho }{{\tilde \sigma }^2}_w} \right)}^\lambda  {{f_{{{\sigma ^2_w}}}}\left( x \right)} dx.
\end{align}
{Differentiating $\bar \xi  \left( \lambda \right)$ with respect to (w.r.t.) $\lambda$, we obtain
\begin{align}\label{adepdd}\
{{2\ln \left( \rho  \right)}}\frac{{\partial \left( {\bar \xi \left( \lambda \right)} \right)}}{{\partial \lambda}} = \left\{ \begin{array}{l}
\begin{array}{*{20}{c}}
{ - \frac{1}{\lambda },}&{}&{\lambda  \le zP_I + \frac{{{{\tilde \sigma }^2}_w}}{\rho }}
\end{array}\\
\begin{array}{*{20}{c}}
{ - \frac{1}{\lambda } + \frac{1}{{\lambda  - zP_I}},}&{}&{\lambda  > zP_I + \frac{{{{\tilde \sigma }^2}_w}}{\rho }}.
\end{array}
\end{array} \right.
\end{align}}
Hence, when ${\lambda  \le zP_I + \frac{{{{\tilde \sigma }^2}_w}}{\rho }}$, $\frac{{\partial \left( {\bar \xi \left( \lambda \right)} \right)}}{{\partial \lambda}}<0$, and when ${\lambda  > zP_I + \frac{{{{\tilde \sigma }^2}_w}}{\rho }}$, $\frac{{\partial \left( {\bar \xi \left( \lambda \right)} \right)}}{{\partial \lambda}}>0$.  Then, according to \eqref{constraint}, to minimize the DEP, the optimum detection threshold $\lambda^*$ is given by
\begin{align}\label{optimalthreshold}\
{\lambda ^*} = \min \left( {zP_I + \frac{{{{\tilde \sigma }^2}_w}}{\rho },\rho {{\tilde \sigma }^2}_w} \right).
\end{align}
Substituting \eqref{optimalthreshold} into \eqref{adep},  we obtain the minimum DEP conditioned on $z$ as
\begin{align}\label{givenz}\
\bar \xi  \left( \lambda \right) {\rm{ = }}\left\{ \begin{array}{l}
{\rm{1 - }}\frac{1}{{{\rm{2}}\ln \left( \rho  \right)}}\ln \left( {1 + \rho z\frac{P_I}{{{{\tilde \sigma }^2}_w}}} \right),z \le \frac{{{{\tilde \sigma }^2}_w}}{P_I}\left( {\rho  - \frac{1}{\rho }} \right)\\
\begin{array}{*{20}{c}}
0&{}&{}&{\begin{array}{*{20}{c}}
{\begin{array}{*{20}{c}}
{\begin{array}{*{20}{c}}
{}&{}
\end{array}}&{}
\end{array}}&{}&{,z > \frac{{{{\tilde \sigma }^2}_w}}{P_I}\left( {\rho  - \frac{1}{\rho }} \right)}.
\end{array}}
\end{array}
\end{array} \right.
\end{align}
Then, averaging $\bar \xi  \left( \lambda \right)$ w.r.t. $z$, {we can obtain the average minimum DEP as
\begin{align}\label{averagexi1}\
\xi  &= \int_0^{\frac{{{{\tilde \sigma }^2}_w}}{P_I}\left( {\rho  - \frac{1}{\rho }} \right)} {\bar\xi \left( \lambda \right)} {f_{{z}}}\left( z \right)dz \nonumber \\
&{{ = }}\int_{\rm{0}}^{\frac{{{{\tilde \sigma }^2}_w}}{{{P_I}}}\left( {\rho  - \frac{1}{\rho }} \right)} {\left( {1 - \frac{1}{{2\ln \left( \rho  \right)}}\ln \left( {1 + \rho z\frac{{{P_I}}}{{{{\tilde \sigma }^2}_w}}} \right)} \right)} \frac{{\exp \left( { - \frac{z}{{{\sigma ^2}_{s,w} + N{\sigma ^2}_{a,s}{\sigma ^2}_{s,w}}}} \right)}}{{\left( {{\sigma ^2}_{s,w} + N{\sigma ^2}_{a,s}{\sigma ^2}_{s,w}} \right)}}dz \nonumber \\
&= 1 - \frac{1}{{2\ln \left( \rho  \right)}}\int_0^{\frac{{{{\tilde \sigma }^2}_w}}{{{P_I}}}\left( {\rho  - \frac{1}{\rho }} \right)} {\exp \left( { - \frac{z}{{{\sigma ^2}_{s,w} + N{\sigma ^2}_{a,s}{\sigma ^2}_{s,w}}}} \right)\frac{{\rho \frac{{{P_I}}}{{{{\tilde \sigma }^2}_w}}}}{{1 + \rho z\frac{{{P_I}}}{{{{\tilde \sigma }^2}_w}}}}} dz\nonumber \\
&= 1 - \frac{1}{{2\ln \left( \rho  \right)}}\int_1^{{\rho ^2}} {\exp \left( { - \frac{{\left( {y - 1} \right)}}{{{{\bar \gamma }^I}_w\rho }}} \right)\frac{1}{y}} dy\nonumber \\
&= 1 - \frac{{\exp \left( {\frac{1}{{{{\bar \gamma }^I}_w\rho }}} \right)}}{{2\ln \left( \rho  \right)}}\left( {\int_{ - \infty }^{{\rho ^2}} {\exp \left( { - \frac{y}{{{{\bar \gamma }^I}_w\rho }}} \right)\frac{1}{y}} dy - \int_{ - \infty }^1 {\exp \left( { - \frac{y}{{{{\bar \gamma }^I}_w\rho }}} \right)\frac{1}{y}} dy} \right)\nonumber \\
&= 1 - \frac{{\exp \left( {\frac{1}{{\rho \bar{\gamma}^I_{w} }}} \right)}}{{2\ln \left( \rho  \right)}} \left( {Ei\left( { - \frac{{\rho}}{{\bar{\gamma}^I_{w}}}} \right) - Ei\left( { - \frac{1}{{\rho \bar{\gamma}^I_{w}}}} \right)} \right)
\end{align}
where $\bar{\gamma}^I_{w}={\frac{{\left( {{\sigma ^2}_{s,w}{\rm{ + }}N{\sigma ^2}_{a,s}{\sigma ^2}_{s,w}} \right)P_I}}{{{{\tilde \sigma }^2}_w}}}$ denotes the average SNR at Willie with IRS assistance, $y = 1 + \rho z\frac{{{P_I}}}{{{{\tilde \sigma }^2}_w}}$, $Ei\left( x \right) = \int_{ - \infty }^x {\exp \left( t \right){t^{ - 1}}} dt$, and ${f_{\rm{z}}}\left( z \right)$ denotes the PDF of $z$, which is derived in Appendix A.}  According to \eqref{singleproblemc} and \eqref{averagexi1}, the maximum allowed transmission power $\tilde{P_I}$ can be computed numerically or using a look-up table at Matlab. In addition, considering the transmission power constraint in \eqref{singleproblemb}, the optimum transmission power is given by $P^*_I=\min\left({P_{max}, \tilde{P_I}}\right)$. After the optimal transmission power is obtained, we derive the optimal phase shifts by maximizing Bob's link power gain, given by ${\left| {{h^H_{a,b}} + {{\bf{g}}^H_{s,b}}{\bf{\Theta}}{{\bf{h}}_{a,s}}} \right|^2}$.  Let ${{\bf{v}}^H} = \left[ {{e^{j{\theta _1}}},{e^{j{\theta _2}}}, \cdots ,{e^{j{\theta _N}}}} \right]$, then, ${{\bf{g}}^H_{s,b}}{\bf{\Theta}}{{\bf{h}}_{a,s}} = {{\bf{v}}^H} diag \left( {{{\bf{g}}^H_{s,b}}} \right){{\bf{h}}_{a,s}}$. Thus,  Bob's link power gain satisfies the following inequality
\begin{align}\label{ineq}\
{\left| {{h^H_{a,b}} + {{\bf{v}}^H}diag\left( {{{\bf{g}}^H_{s,b}}} \right){{\bf{h}}_{a,s}}} \right|^2}\leq \left( \left|{h^H_{a,b}}\right|+ \left|{{{\bf{v}}^H}diag\left( {{{\bf{g}}^H_{s,b}}} \right){{\bf{h}}_{a,s}}} \right|\right)^2.
 \end{align}
From \eqref{ineq}, we find that when equality holds,  ${h^H_{a,b}}$ and ${{\bf{v}}^H}diag\left( {{{\bf{g}}^H_{s,b}}} \right){{\bf{h}}_{a,s}}$ have the same angle, and Bob's link power gain is maximized. Thus, the optimal IRS phase shifts are given by
 \begin{align}\label{averagexi}\
 {\theta^* _i} = arg \left(h_{a,b}\right)-{arg({{g}^H_{{s_i},b}})}-{arg({h_{a,{s_i}}})}, 1\leq i\leq N.
 \end{align}
 At the optimal ${\theta^* _i}$, the SNR at Bob is given by
 \begin{align}\label{rb}\
{\gamma'}_{IRS} = \frac{{{P^*_I}{{\left( {\sum\limits_{i = 1}^N {\left| {{h_{a,{s_i}}}} \right|\left| {{g_{{s_i},b}}} \right|}  + \left| {{h_{a,b}}} \right|} \right)}^2}}}{{{\sigma^2_b}}}.
 \end{align}

\emph{Discussion:} For comparison, we study the covert transmission rate without IRS assistance. For this case,  $z$ is reduced to $z={{\left| {{h_{a,w}}} \right|}^2}$, and the PDF of $z$ is ${f_z}\left( z \right) = \frac{1}{{{\sigma ^2}_{a,w}}}\exp \left( { - \frac{z}{{{\sigma ^2}_{a,w}}}} \right)$. Thus, the DEP without IRS aid is given by
\begin{align}\label{averagexi2}\
\xi &= 1 - \frac{{\exp \left( {\frac{1}{{\rho \bar\gamma^d_{D}}}} \right)}}{{2\ln \left( \rho  \right)}} \left( {Ei\left( { - \frac{{\rho}}{\bar\gamma^d_{D}}} \right) - Ei\left( { - \frac{{1}}{{\rho \bar\gamma^d_{D}}}} \right)} \right)
\end{align}
where $\bar\gamma^d_{D}={\frac{{{{\sigma ^2}_{s,w}} P_D}}{{{{\tilde \sigma }^2}_w}}}$ denotes the average SNR at Willie without IRS assistance.  According to the covert constraint and \eqref{averagexi2}, the allowed transmission power $\tilde{P_D}$ can be calculated. Then, considering the maximum power constraint,  the optimal direct transmission power is $
 P^*_D=\min\left({P_{max}, \tilde{P_D}}\right)$. Thus, without IRS assistance, the SNR at Bob is given by
  \begin{align}\label{rbd}\
 {\gamma'}_{dir} = \frac{{P^*_D{{\left( {\left| {{h_{a,b}}} \right|} \right)}^2}}}{{{\sigma^2_b}}}.
  \end{align}
 Compared  \eqref{rb} to \eqref{rbd}, we find that the IRS strengthens Bob's link power gain since \\ ${{\left( {\sum\limits_{i = 1}^N {\left| {{h_{a,{s_i}}}} \right|\left| {{g_{{s_i},b}}} \right|}  + \left| {{h_{a,b}}} \right|} \right)}^2}\geq \left| {{h_{a,b}}} \right|^2$. However, the signal power received at Willie is also strengthened due to IRS assistance, which can be seen from \eqref{pdfz}. As a result,  $P^*_D \geq P^*_I$.  Hence, when only partial CSI of Willie's link is available, it is not necessary that the covert rate of IRS aided system is greater than that without IRS assistance. Only when the covert constraint is loose, or  Willie's link quality is poor, does IRS become beneficial to covert transmission performance. In these cases, the covert constraint has little impact on the transmission power at Alice, and the effect of Bob's link dominates the covert transmission rate.
\subsection{With Instantaneous CSI of Willie's Link}
{In practice, Alice, Bob and Willie can be in the same network, where Alice wishes to send
confidential messages to Bob, and assumes Willie to be an untrusted or suspicious node. For this
scenario, Willie can be friendly to Alice, and Alice has perfect knowledge of the instantaneous
CSI of Willie's link. When Willie is friendly to Alice, several methods can be used to estimate the IRS link. The active channel sensors can be inserted into the array of IRS passive elements to sense channel information \cite{ATaha}. In addition, the active sensors can work in the channel sensing mode for channel estimation and in the reflection mode for electromagnetic reflection.  This two-function method at active sensors increases the hardware complexity and requires extra energy and time to transfer the estimated channel information towards the controller. Moreover, the cascaded IRS channels can be decomposed into a series of sub-channels, which can be realized by turning on one IRS element and turning off the other IRS elements \cite{DMishra}. Finally, structure-learning based CSI acquisition can also be used by exploiting  strong structural features, such as sparsity and low-rank. With this structural information, the estimation of a cascaded channel can be performed by utilizing advanced signal processing tools such as
compressed sensing, sparse matrix factorization, and low-rank matrix recovery algorithms \cite{ZQHe,ZWang,XYuan}}. Moreover, the covert rate with perfect instantaneous CSI of Willie's link can be an upper bound on that with partial or imperfect CSI of Willie's link. Hence, it is necessary to investigate the covert rate with perfect CSI of Willie's link. In this case, the transmission power at Alice and IRS phase shifts are changed continuously according to both Bob's link and Willie's link, which is different from the partial CSI case. To be undetected by Willie, the DEP should be larger than $1-\kappa$.  According to \eqref{givenz}, the signal power received at Willie should be less than the covert transmission threshold $\eta  = \min \left( {{{\tilde \sigma }^2}_w\left( {\rho  - \frac{1}{\rho }} \right),\frac{{\left( {\rho^{2\kappa }  - 1} \right){{\tilde \sigma }^2}_w}}{\rho }} \right)$. Therefore, the optimization problem for covert rate maximization is given by
\begin{align}\label{optimalsinglewith}\
&\mathop {\max }\limits_{{\bf{v}}, P_I} \quad {P_I}{\left| {{h^H}_{a,b} + {{\bf{v}}^H} diag \left( {{{\bf{g}}^H_{s,b}}} \right){{\bf{h}}_{a,s}}} \right|^2}\nonumber \\
&s.t.\quad {P_I} \le {P_{\max }}\nonumber \\
&\quad\quad {{P_I}{{\left| {{h^H}_{a,w} + {{\bf{v}}^H} diag \left( {{{\bf{g}}^H_{s,w}}} \right){{\bf{h}}_{a,s}}} \right|}^2} \le \eta }
\end{align}
where the transmission power $P_I$ and phase shifts vector $\bf{v}$ are coupled. It is challenging to derive the exact expressions for the optimal $P_I$ and ${\bf{v}}$. Hence, we use an alterative algorithm to solve this optimization problem. For a given phase shifts vector ${\bf{v}}$, to satisfy the covert constraint, the optimal transmission power is given by
 \begin{align}\label{optimalpower}\
P^*_I = \min \left( {{P_{\max }},\frac{\eta }{{{{\left| {{h^H}_{a,w} + {{\bf{v}}^H} diag \left( {{{\bf{g}}^H_{s,w}}} \right){{\bf{h}}_{a,s}}} \right|}^2}}}} \right).
 \end{align}
Then, for a given $P_I$, the optimization problem \eqref{optimalsinglewith} is rewritten as
\begin{align}\label{optimalsinglewith21}\
&\mathop {\max }\limits_{\bf{V}}\quad {P_I}\left( {Tr\left( {{{\bf{T}}{_{a,b}\bf{V}}}} \right) + {{\left| {{h_{a,b}}} \right|}^2}} \right)\nonumber \\
&{s.t.}\quad {{P_{{I}}}\left( {Tr\left( {{\bf{T}}_{a,w}{\bf{V}}} \right) + {{\left| {{h_{a,w}}} \right|}^2}} \right) \le \eta }\nonumber \\
&\quad\quad\quad {{{\bf{V}}_{n,n}} = 1},{{\bf{V}} \succeq 0}, {rank}\left( {\bf{V}} \right) = 1
   \end{align}
 where  ${\bf{V}} = {\bf{\bar v}}{{{\bf{\bar v}}}^H}$, ${{{\bf{\bar v}}}^H} = t^H\left[ {{{\bf{v}}^H},1} \right]$, and $t$ is an
auxiliary variable satisfying $\left| t \right| = 1$. In addition, ${{\bf{T}}_{a,j}}$ $ (j\in (b,w) )$  is  given by
 \begin{align}\label{Qmatrix}\
{{\bf{T}}_{a,j}}{{ = }}\left[ \begin{array}{l}
diag\left( {{{\bf{g}}^H}_{s,j}} \right){{\bf{h}}_{a,s}}{{\bf{h}}^H}_{a,s}diag\left( {{{\bf{g}}_{s,j}}} \right),diag\left( {{{\bf{g}}^H}_{s,j}} \right){{\bf{h}}_{a,s}}{{{h}}_{a,j}}\\
{{{h}}^H}_{a,j}{{\bf{h}}^H}_{a,s}diag\left( {{{\bf{g}}_{s,j}}} \right),{\bf{0}}
\end{array} \right].
 \end{align}
The optimization problem in \eqref{optimalsinglewith21} is non-convex due to the constraint of $rank({\bf{V}})=1$. { Successive convex approximation (SCA) solves the second-order cone programming (SOCP) problem by replacing the non-convex constraint with a convex constraint \cite{LTran,Hhan}. However, it is challenging to find a convex restriction to replace the non-convex constraint in \eqref{optimalsinglewith}. By contrast, SDR can be used to solve the SOCP by relaxing the non-convex constraint. Hence, we use SDR to solve the optimization problem.} We omit the constraint $rank({\bf{V}})=1$, and write the optimization problem as follows
\begin{align}\label{optimalsinglewith2}\
&\mathop {\max }\limits_{\bf{V}}\quad {P_I}\left( {Tr\left( {{{\bf{T}}{_{a,b}\bf{V}}}} \right) + {{\left| {{h_{a,b}}} \right|}^2}} \right)\nonumber \\
&{s.t.}\quad {{P_{{I}}}\left( {Tr\left( {{\bf{T}}_{a,w}{\bf{V}}} \right) + {{\left| {{h_{a,w}}} \right|}^2}} \right) \le \eta }\nonumber \\
&\quad\quad\quad {{{\bf{V}}_{n,n}} = 1},{{\bf{V}} \succeq 0}.
   \end{align}
It is clear that \eqref{optimalsinglewith2} is a convex problem and can be solved by the SDP technique.
The optimization tool, such as CVX in Matlab, can be used to compute the optimal ${\bf{V}}^*$. Then, the Gaussian randomization technique is used to calculate the optimal IRS phase shifts vector satisfying the constraint of $rank\left({{\bf{V}}^*}\right)=1$. In Gaussian randomization, we first calculate the eigen-decomposition of ${\bf{V}}^*={\bf{X}\Sigma\bf{X}}^H$ and make ${\bf{\bar{v'}}}_l=\arg\left( {\bf{X}}{\bf{\Sigma}}^{1/2}{\bf{e}}_l\right)$, where ${\bf{e}}_l (1\leq l \leq L )$ is a vector of zero-mean, unit-variance complex circularly Gaussian random variables, and $L$ is the number of randomizations. Let ${\bf{V}}'_l={\bf{\bar{v'}}}_l{\bf{\bar{v'}}}^H_l$, and search the ${\bf{V}}'_l$ achieving the largest covert transmission rate as well as satisfying the covert constraint in \eqref{optimalsinglewith2}. Finally, according to the corresponding ${\bf{\bar{v}}}'_l$, the optimal $\bf{v}^*$ is computed as ${\bf{v}^*}=\frac{{[{\bf{\bar{v}}}'_l]}_{\left(1:N\right)}}{{{\bf{\bar{v}}}'_l}\left(N+1\right)}$, where $[{\bf{\bar{v}}}'_l]_{(1:N)}$ denotes the elements of Row $1$ to Row $N$ in ${\bf{\bar{v}}}'_l$. Then, $P^*_I$ and ${\bf{v}^*}$ are iteratively calculated until the desired accuracy of covert transmission is satisfied.

For this alterative algorithm, assume that at the end of the $t$-th iteration, the calculated optimal power and phase shifts vector are ${{P_I}}^t$ and ${\bf{v}}^t$. Then, in the $t+1$  iteration, we can calculate the optimal ${{P_I}}^{t+1}$ given ${\bf{v}}^t$ and the optimal ${\bf{v}}^{t+1}$ given ${{P_I}}^{t+1}$. Assume that the objective function is $f\left({{P_I}}, {\bf{v}} \right)$. It is obvious that $f\left( {{P^t}_I,{{\bf{v}}^t}} \right) \le f\left( {{P^{t + 1}}_I,{{\bf{v}}^t}} \right) \le f\left( {{P^{t + 1}}_I,{{\bf{v}}^{t + 1}}} \right)$. This guarantees that the alternating algorithm is non-decreasing. Since there are $N+1$ variables and 1 linear constraint in the optimization problem \eqref{optimalsinglewith2},  the complexity of calculating the phase shifts vector is $O\left( {{{\left( {{{\left( {N + 1} \right)}^{{2}}}{{ + 1}}} \right)}^{{{3}}{{.5}}}}} \right)$ for each iteration \cite{ZhiQuanLuo}. In addition, the complexity of calculating the transmission power can be neglected according to \eqref{optimalpower}. Hence, using this alternating algorithm, the complexity of \eqref{optimalsinglewith} is $O\left( {{{\left( {{{\left( {N + 1} \right)}^{{2}}}{{ + 1}}} \right)}^{{{3}}{{.5}}}}} \right)$ for each iteration. {To compare the covert rate with IRS assistance to that with direct transmission, we need to derive the SNR at Bob under the covert constraint. When IRS is used to forward the messages, it is challenging to derive a closed-form expression for SNR at Bob. Hence, with IRS assistance, we alternatively investigate an upper bound on the SNR at Bob. It is noted that a lower-bound on Willie's link power gain is
 \begin{align}\label{Qmatrix}\
{\left| {{h^H_{a,w}} + {{\bf{g}}^H_{s,w}}\bf{\Theta}}{{\bf{h}}_{a,s}} \right|^2}& = {{{\bf{\bar v}}}^H}{{\bf{T}}_{a,w}}{\bf{\bar v}} + {\left| {{h^H}_{a,w}} \right|^2}{\geq}{\lambda _{\min }}\left( {{{\bf{T}}_{a,w}}} \right)\left( {{\rm{N + 1}}} \right){\rm{ + }}\left| {{h^H}_{a,w}} \right|^2
  \end{align}
 where ${\lambda _{\min }}\left( {{{\bf{T}}_{a,w}}} \right)$ is the minimum eigenvalue of ${{{\bf{T}}_{a,w}}}$, and the equality in the right side of (25) holds when ${\bf{\bar v}}$ is the eigenvector corresponding to the minimum eigenvalue ${\lambda _{\min }}\left( {{{\bf{T}}_{a,w}}} \right)$. In addition, an upper bound on Bob's link power gain is given by
\begin{align}\label{Qmatrix}\
{\left| {{h^H_{a,b}} + {{\bf{g}}^H_{s,b}}\bf{\Theta}}{{\bf{h}}_{a,s}} \right|^2}\le {{{{\left( {\left| {{h_{a,b}}} \right| + \sum\limits_{k = 1}^N {\left| {{h_{a,{s_k}}}} \right|} \left| {{g_{{s_k},b}}} \right|} \right)}^2}}}
\end{align}
where the equality holds when $\arg(h^H_{a,b})$ is equal to $\arg ({{\bf{g}}^H_{s,b}}\bf{\Theta}{{\bf{h}}_{a,s}})$. Thus, we can obtain the following upper bound on the SNR at Bob
\begin{align}\label{gammaIRSu}\
{\gamma ^u}_{IRS} = \frac{{\eta \left( {\left| {{h_{a,b}}} \right| + \sum\limits_{k = 1}^N {\left| {{h_{a,{s_k}}}} \right|} \left| {{g_{{s_k},b}}} \right|} \right)^2}}{\sigma^2_{b}{\left({\lambda _{\min }}\left( {{{\bf{T}}_{a,w}}} \right)\left( {{\rm{N + 1}}} \right){\rm{ + }}\left| {{h^H}_{a,w}} \right|^2\right)}}.
\end{align}
By contrast,  the SNR at Bob without IRS is given by ${\gamma _{dir}} = \frac{{\eta {{\left| {{{\rm{h}}^H}_{a,b}} \right|}^{\rm{2}}}}}{\sigma^2_{b}{{{\left| {{h_{a,w}}} \right|}^2}}}$.} Then, if ${\gamma ^u}_{IRS} <{\gamma _{dir}}$, the covert transmission rate without IRS is greater than that with IRS.  Note that if ${\lambda _{\min }}\left( {{{\bf{T}}_{a,w}}} \right) > 0$, compared to the direct transmission, Willie's link power gain with IRS is strengthened. To satisfy the covert constraint, the transmission power is reduced compared to direct transmission. As a result, the covert transmission rate without IRS can outperform that with IRS.  By contrast, if ${\lambda _{\min }}\left( {{{\bf{T}}_{a,w}}} \right) < 0$,  the covert transmission rate with IRS is greater than that without IRS. Hence, it is not necessary that the IRS is beneficial to the covert transmission rate for the single transmit antenna case. Many factors, such as the predetermined covert threshold, the relationship between Willie's link power gain
and Bob's link power gain, and the number of IRS elements, determine whether the transmission with IRS outperforms that without IRS or not, even if the full instantaneous CSIs of all links are perfectly known at Alice.
\section{Multiple Antennas at Alice}
When there are multiple antennas at Alice, transmit beamforming technology can be used to achieve a higher covert performance. In this section, we investigate  joint transmit beamforming and phase shift optimization for covert transmission rate maximization. Next, we discuss the two cases of partial CSI and instantaneous CSI of Willie's link.
\subsection{With Partial CSI of  Willie's Link}
{In the case of partial CSI of Willie's link, only the statistical information, i.e., mean and variance, of the Alice-to-Willie and IRS-to-Willie channels are available, and the transmit beamforming vector at Alice and phase shifts vector at the IRS are jointly optimized  and updated according to the instantaneous CSI of Bob's link.}  Thus, the transmit beamforming and phase shifts vectors are random for Willie's link. For the case of random beamforming towards Willie, the received SNR at Willie has the same expression as that with a single antenna. As a result, the maximum transmission power with multiple antennas at Alice has the same expression as that with a single antenna at Alice. After the transmission power is determined, the transmit beamforming and the phase shifts vectors, which are coupled, are jointly optimized for the covert rate maximization using an alterative algorithm. For a given phase shifts vector ${\bf{v}}$, the optimal beamforming vector for MISO systems is MRT, i.e. ${\bf{{\bf{w}}} ^{{*}}}{\rm{ = }}\sqrt{P_I}\frac{{{{\left( {{\bf{h}^H}_{a,b} + {{\bf{v}}^H} diag \left( {{{\bf{g}}^H_{s,b}}} \right){\bf{h}}_{a,s}} \right)}^H}}}{{\left\| {{\bf{h}^H}_{a,b} + {{\bf{v}}^H} diag \left( {{{\bf{g}}^H_{s,b}}} \right){\bf{h}}_{a,s}} \right\|}}\in \mathbb{C}^{M\times 1}$. When ${\bf{{\bf{w}}} ^{{*}}}$  and ${P^*_I}$ are determined, the optimization problem is reduced to maximizing Bob's link power gain, and it is given by
\begin{align}\label{optimalproblem1}\
&\mathop {\max }\limits_{\bf{v}}  {\left\| {{{\left( {{{\bf{h}}^H}_{a,b} +{{\bf{v}}^H} diag \left( {{{\bf{g}}^H_{s,b}}} \right){\bf{h}}_{a,s}} \right)}}} \right\|^2}=\mathop {\max }\limits_{\bf{v}}  \left({{\bf{v}}^H}{ diag\left( {{{\bf{g}}^H}_{s,b}} \right){{\bf{h}}_{a,s}}}{{{\bf{h}}^H_{a,s}}diag\left( {{{\bf{g}}}_{s,b}} \right)}{\bf{v}} \right. \nonumber \\
 &\quad\quad\quad \left.+ {{\bf{h}}^H}_{a,b}{{{{\bf{h}}^H_{a,s}}diag\left( {{{\bf{g}}}_{s,b}} \right)}}{\bf{v}}+ {{\bf{v}}^H}{ diag\left( {{{\bf{g}}^H}_{s,b}} \right){{\bf{h}}_{a,s}}}{{\bf{h}}_{a,b}} + {\left\| {{{\bf{h}}^H}_{a,b}} \right\|^2}\right).
 \end{align}
 Similar to \eqref{optimalsinglewith2}, this optimization problem can be rewritten as
 \begin{subequations}
\begin{align}\label{optimalV}\
&{\mathop {\max }\limits_{\bf{V}} }\quad {Tr\left( {{\bf{R_{a,b}V}}} \right)+  {\left\| {{{\bf{h}}^H}_{a,b}} \right\|^2}}\\
&{s.t.}\quad{{{\bf{V}}_{n,n}} = 1,}n = 1, \cdots N + 1 \\
&{{\bf{V}} \succeq 0},{Rank}\left( {\bf{V}} \right) = 1
 \end{align}
 \end{subequations}
where ${{\bf{R}}_{a,b}} = \left[
{{\bf{\Phi }}_{a,b}}{{\bf{\Phi }}^H}_{a,b},{{\bf{\Phi}}_{a,b}}{\alpha^H _{a,b}};
{\alpha _{a,b}}{{\bf{\Phi }}^H}_{a,b},\bf{0}\right]$,
${{\bf{\alpha}} _{a,b}} = {{\bf{h}}^H}_{a,b}{\bf{{\bf{w}} }}$, and ${{\bf{\Phi }}_{a,b}} = diag\left( {{{\bf{g}}^H}_{s,b}} \right){{\bf{h}}_{a,s}}{\bf{{\bf{w}} }}$. The optimal ${\bf{v}}^*$ in \eqref{optimalproblem1} can be calculated using the CVX and Gaussian randomization \cite{Qwu1}, and the maximum covert transmission rate is obtained by substituting ${\bf{v}}^*$ into \eqref{optimalproblem1}. For this case, no iteration is required.  But, it is challenging to derive an exact SNR expression at Bob. Alternatively, we obtain the following upper bound on Bob's link power gain
\begin{align}\label{SNRatBob1}\
&{\mathbb{E}}\left( {{{\left\| {{{\bf{h}}^H}_{a,b} + {{\bf{v}}^H} diag \left( {{{\bf{g}}^H_{s,b}}} \right){\bf{h}}_{a,s}} \right\|}^2}} \right)\nonumber \\
 &\le M\left( {{\sigma ^2}_{a,b} + N\left( {1 + \frac{{\left( {N - 1} \right){\pi ^2}}}{{16}}} \right){\sigma ^2}_{a,s}{\sigma ^2}_{s,b} + \frac{{N{\pi ^{\frac{3}{2}}}{\sigma _{a,b}}{\sigma _{a,s}}{\sigma _{s,b}}}}{4}} \right).
 \end{align}
The proof of \eqref{SNRatBob1} can be found in \cite{EBasar,Hhan}. {In \eqref{SNRatBob1}, equality holds when $N \gg 1$, and \\
$\arg({{\bf{h}}^H}_{a,b})=\arg({{\bf{v}}^H} diag \left( {{{\bf{g}}^H_{s,b}}} \right){\bf{h}}_{a,s})$}.
\subsection{With Instantaneous CSI of  Willie's Link}
When the instantaneous CSI of Willie's link is perfectly known at Alice, to maximize the covert transmission rate, the transmit beamforming and phase shifts vectors are continuously adjusted according to the instantaneous CSIs of both Willie's and Bob's links. In this case, the optimization problem is formulated as follows
\begin{align}\label{problem2}\
&\mathop {\max }\limits_{\bf{v} ,{\bf{w}} } {\left| {\left( {{{\bf{h}}^H}_{a,b} + {{\bf{v}}^H} diag \left( {{{\bf{g}}^H_{s,b}}} \right){\bf{h}}_{a,s}} \right)\bf{{\bf{w}}} } \right|^2}\nonumber \\
& s.t.\quad {Tr\left( {{\bf{w}} {{\bf{w}} ^H}} \right)} \le {P_{\max }}\nonumber \\
&{{{\left| {\left( {{{\bf{h}}^H}_{a,w} + {{\bf{v}}^H} diag \left( {{{\bf{g}}^H_{s,w}}} \right){\bf{h}}_{a,s}} \right)\bf{{\bf{w}}} } \right|}^2} \le \eta }.
 \end{align}
 For this optimization problem, we first present an alternating algorithm to maximize the covert rate. Then, we propose two suboptimal algorithms to reduce the complexity. The details of these algorithms are discussed next.
 \subsubsection{Alternating algorithm}
 For a given phase shifts vector $\bf{v}$, we need to optimize the beamforming vector $\bf{{\bf{w}}}$. For the MISO system with covert constraint, we can resort to the traditional MISO cognitive system, which has the interference constraint at the primary user and has a similar optimization objective function. Thus, by leveraging results from MISO cognitive systems in \cite{RZhang,YPei}, we can derive the optimal beamforming vector as ${\bf{{\bf{w}}^*}}=P_{max}{{\bf{\bar{\bf{w}} }}} $, where  ${{\bf{\bar{\bf{w}} }}}$ denotes the unit-norm beamforming vector and is given by
\begin{align}\label{optimalomega}\
&{{\bf{\bar{\bf{w}} }}} = A\frac{{\left( {{{\bf{h}}^H}_{a,b} + {{\bf{v}}^H} diag \left( {{{\bf{g}}^H_{s,b}}} \right){\bf{h}}_{a,s}} \right)^H}}{{\left\| {{{\bf{h}}^H}_{a,b} + {{\bf{v}}^H} diag \left( {{{\bf{g}}^H_{s,b}}} \right){\bf{h}}_{a,s}} \right\|}}\nonumber \\
 &+ B\frac{{\left( {{\bf{I}}_M - \frac{{\left( {{{\bf{h}}^H}_{a,w} + {{\bf{v}}^H} diag \left( {{{\bf{g}}^H_{s,w}}} \right){\bf{h}}_{a,s}} \right)^H{{\left( {{{\bf{h}}^H}_{a,w} + {{\bf{v}}^H} diag \left( {{{\bf{g}}^H_{s,w}}} \right){\bf{h}}_{a,s}} \right)}}}}{{{{\left\| {{{\bf{h}}^H}_{a,w} + {{\bf{v}}^H} diag \left( {{{\bf{g}}^H_{s,w}}} \right){\bf{h}}_{a,s}} \right\|}^2}}}} \right)\left( {{{\bf{h}}^H}_{a,b} + {{\bf{v}}^H} diag \left( {{{\bf{g}}^H_{s,b}}} \right){\bf{h}}_{a,s}} \right)^H}}{{\left\| {\left( {{\bf{I}}_M - \frac{{\left( {{{\bf{h}}^H}_{a,w} + {{\bf{v}}^H} diag \left( {{{\bf{g}}^H_{s,w}}} \right){\bf{h}}_{a,s}} \right)^H{{\left( {{{\bf{h}}^H}_{a,w} + {{\bf{v}}^H} diag \left( {{{\bf{g}}^H_{s,w}}} \right){\bf{h}}_{a,s}} \right)}}}}{{{{\left\| {{{\bf{h}}^H}_{a,w} + {{\bf{v}}^H} diag \left( {{{\bf{g}}^H_{s,w}}} \right){\bf{h}}_{a,s}} \right\|}^2}}}} \right){{\left( {{{\bf{h}}^H}_{a,b} + {{\bf{v}}^H} diag \left( {{{\bf{g}}^H_{s,b}}} \right){\bf{h}}_{a,s}} \right)}^H}} \right\|}}
 \end{align}
where
\begin{align}\label{raw}\
\left( {A,B} \right) = \left\{ \begin{array}{l}
\left( {1,0} \right),\eta  \ge {P_{\max }}\left( {\left\| {{{\bf{h}}^H}_{a,w} + {{\bf{v}}^H} diag \left( {{{\bf{g}}^H_{s,w}}} \right){\bf{h}}_{a,s}} \right\|\cos {\Omega} } \right)^2\\
\left( {\frac{\tau}{{\cos {\Omega} }},\sqrt {1 - {\tau ^2}}  - \tau \tan {\Omega} } \right), otherwise
\end{array} \right.
 \end{align}
 $\tau = \sqrt {\frac{\eta }{{{P_{\max }}{{\left\| {{{\bf{h}}^H}_{a,w} + {{\bf{v}}^H} diag \left( {{{\bf{g}}^H_{s,w}}} \right){\bf{h}}_{a,s}} \right\|}^2}}}}$, and $\cos {\Omega}  = \frac{{\left( {{{\bf{h}}^H}_{a,w} + {{\bf{v}}^H} diag \left( {{{\bf{g}}^H_{s,w}}} \right){\bf{h}}_{a,s}} \right){{\left( {{{\bf{h}}^H}_{a,b} +{{\bf{v}}^H} diag \left( {{{\bf{g}}^H_{s,b}}} \right){\bf{h}}_{a,s}} \right)}^H}}}{{\left\| {{{\bf{h}}^H}_{a,w} + {{\bf{v}}^H} diag \left( {{{\bf{g}}^H_{s,w}}} \right){\bf{h}}_{a,s}} \right\|\left\| {{{\bf{h}}^H}_{a,b} + {{\bf{v}}^H} diag \left( {{{\bf{g}}^H_{s,b}}} \right){\bf{h}}_{a,s}} \right\|}}$. For a given $\bf{{\bf{w}}}$ , {similar to \eqref{optimalsinglewith2}}, the optimization problem \eqref{problem2} is rewritten as
 \begin{align}\label{perproblem3}\
&{\mathop {\max }\limits_{\bf{v}} }\quad {{{\left| {{\alpha _{a,b}}} \right|}^2} + Tr\left( {{{\bf{R}}_{a,b}}{\bf{V}}} \right)}
\nonumber \\
&{s.t.}\quad\quad  {{{\left| {{\beta _{a,w}}} \right|}^2} + Tr\left( {{{\bf{R}}_{a,w}}{\bf{V}}} \right) \le \eta }
\nonumber\\
&\quad\quad {{{\bf{V}}_{n,n}} = 1}, {{\bf{V}} \succeq 0}, rank\left(  {\bf{V}}\right)=1
 \end{align}
where${{\bf{R}}_{a,w}} = \left[
{{\bf{\Psi }}_{a,w}}{{\bf{\Psi }}^H}_{a,w},{{\bf{\Psi }}_{a,w}}{\beta^H _{a,w}};
{\beta _{a,w}}{{\bf{\Psi }}^H}_{a,w}, \bf{0} \right]$,
 ${\beta _{a,w}} = {{\bf{h}}^H}_{a,w}{\bf{{\bf{w}} }}$, and \\
 ${{\bf{\Psi }}_{a,w}} = diag\left( {{{\bf{g}}^H}_{s,w}} \right){{\bf{h}}_{a,s}}{\bf{{\bf{w}} }}$. When neglecting the constraint $rank\left(  {\bf{V}}\right)=1$,  problem \eqref{perproblem3} can be  solved using the software package CVX.  Then, Gaussian randomization can be used to construct a rank-one solution. Finally, iterations for transmit beamforming and phase shift vectors optimization are performed for covert rate maximization. The complexity of solving \eqref{perproblem3} is $O\left( {{{\left( {{{\left( {N + 1} \right)}^{\rm{2}}}{\rm{ + 1}}} \right)}^{{\rm{3}}{\rm{.5}}}}} \right)$ for each iteration. In the alternating algorithm, for a given beamforming vector $\bf{w}$, it is possible that there is no available phase shifts vector satisfying the covert constraint. Hence, we first obtain the optimal $\bf{w}$ for a given phase shifts vector. The order for the alterative algorithm cannot be changed. In practice, the number of iterations is determined by the covert transmission rate accuracy. For example, when the gap between the covert transmission rate for the $(i-1)$-th iteration and for the $i$-th iteration is less than a predetermined threshold $\Gamma$, the alterative algorithm ends at the $i$-th iteration.

\subsubsection{Zero-forcing Beamforming}
To decrease the complexity of the alterative algorithm, we propose a suboptimal algorithm, where zero-forcing beamforming is used at Alice, and no legitimate signal can be received at Willie. Thus, Alice can transmit messages with the maximum power, namely, $P^*_I=P_{max}$, and the covert constraint is always satisfied. Since it is challenging to first derive the zero-forcing beamforming, and then achieve optimal phase shifts, we give a alterative algorithm with zero-beamforming deployed at Alice. Specifically, for a given beamforming vector $\bf{w}$, the phase shifts vector ${\bf{v}}$ is optimized for covert rate maximization, and the optimization problem is written as
\begin{align}\label{optimalproblemzeroforcing}\
\mathop {\max }\limits_{\bf{v}}  \left| {\left( {{{\bf{h}}^H}_{a,b} + {{\bf{v}}^H} diag \left( {{{\bf{g}}^H_{s,b}}} \right){\bf{h}}_{a,s}} \right){\bf{w}} } \right|^2.
 \end{align}
Since $\left| {\left( {{{\bf{h}}^H}_{a,b} + {{\bf{v}}^H} diag \left( {{{\bf{g}}^H_{s,b}}} \right){\bf{h}}_{a,s}} \right){\bf{w}} } \right| \le \left| {{{\bf{h}}^H}_{a,b}{\bf{w}} } \right| + \left| {{{\bf{v}}^H} diag \left( {{{\bf{g}}^H_{s,b}}} \right){\bf{h}}_{a,s}{\bf{w}} } \right|$, to maximize the covert transmission rate, the following equality should be satisfied
 \begin{align}\label{angleoptimalproblem}\
\arg \left( {{{\bf{h}}^H}_{a,b}{\bf{w}} } \right) = \arg \left( {{{\bf{v}}^H}diag\left( {{{\bf{g}}^H}_{s,b}} \right){{\bf{h}}_{a,s}}{\bf{w}} } \right).
  \end{align}
According to \eqref{angleoptimalproblem}, we can obtain the optimal ${\theta _i}^*$ $\left(  1 \leq i \leq N \right)$ as
 \begin{align}\label{optimalthetazero}\
 {\theta _i}^* =\arg \left( {{{\bf{h}}^H}_{a,b}{\bf{w}} } \right) - \arg \left( {{g^H}_{{s_i},b}} \right) - \arg \left( {{{\bf{h}}}_{a,{s_i}}{\bf{{\bf{w}} }}} \right).
  \end{align}
Conditioned on the derived phase shift ${\theta _i}^*$, the zero-forcing transmit beamforming vector is
\begin{align}\label{zeroforcing}\
{{\bf{{\bf{w}} }}}{{ = }}{P_{max}}\frac{{\left( {{\bf{I}}_M - \frac{{\left( {{{\bf{h}}^H}_{a,w} + {{\bf{v}}^H} diag \left( {{{\bf{g}}^H_{s,w}}} \right){\bf{h}}_{a,s}} \right){{\left( {{{\bf{h}}^H}_{a,w} + {{\bf{v}}^H} diag \left( {{{\bf{g}}^H_{s,w}}} \right){\bf{h}}_{a,s}} \right)}^H}}}{{{{\left\| {{{\bf{h}}^H}_{a,w} + {{\bf{v}}^H} diag \left( {{{\bf{g}}^H_{s,w}}} \right){\bf{h}}_{a,s}} \right\|}^2}}}} \right){{\left( {{{\bf{h}}^H}_{a,b} + {{\bf{v}}^H} diag \left( {{{\bf{g}}^H_{s,b}}} \right){\bf{h}}_{a,s}} \right)}^H}}}{{\left\| {\left( {{\bf{I}}_M - \frac{{\left( {{{\bf{h}}^H}_{a,w} + {{\bf{g}}^H}_{s,w}\Theta {{\bf{h}}_{a,s}}} \right){{\left( {{{\bf{h}}^H}_{a,w} + {{\bf{v}}^H} diag \left( {{{\bf{g}}^H_{s,w}}} \right){\bf{h}}_{a,s}} \right)}^H}}}{{{{\left\| {{{\bf{h}}^H}_{a,w} + {{\bf{v}}^H} diag \left( {{{\bf{g}}^H_{s,w}}} \right){\bf{h}}_{a,s}} \right\|}^2}}}} \right){{\left( {{{\bf{h}}^H}_{a,b} + {{\bf{v}}^H} diag \left( {{{\bf{g}}^H_{s,b}}} \right){\bf{h}}_{a,s}} \right)}^H}} \right\|}}
 \end{align}
where ${{\bf{I}}_M - \frac{{\left( {{{\bf{h}}^H}_{a,w} + {{\bf{v}}^H} diag \left( {{{\bf{g}}^H_{s,w}}} \right){\bf{h}}_{a,s}} \right)^H{{\left( {{{\bf{h}}^H}_{a,w} +{{\bf{v}}^H} diag \left( {{{\bf{g}}^H_{s,w}}} \right){\bf{h}}_{a,s}} \right)}}}}{{{{\left\| {{{\bf{h}}^H}_{a,w} + {{\bf{v}}^H} diag \left( {{{\bf{g}}^H_{s,w}}} \right){\bf{h}}_{a,s}} \right\|}^2}}}}$  is a Hermitian orthogonal projection matrix. We iterate between \eqref{optimalthetazero} and \eqref{zeroforcing} for  covert rate maximization. Since the exact analytical expression of phase shifts and transmit beamforming vectors can be obtained, this algorithm has a low complexity  $O\left( {M + N} \right)$ for each iteration.
\subsubsection{Minimizing Willie's Link Power Gain}
Minimizing the channel quality of Willie's link is an effective way to improve the covert transmission rate. Next, we adjust the IRS phase shifts vector to minimize Willie's link power gain. Since the phase shifts and transmit beamforming vectors are coupled, we solve the optimization problem by the iteration method. For a given phase shifts vector $\bf{v}$,  the optimal beamforming vector $\bf{w}$ can be obtained according to \eqref{optimalomega}. Then, for a given beamforming vector $\bf{w}$, we minimize Willie's link power gain  by adjusting the phase shifts vector $\bf{v}$. Note that the signal power received at Willie is expressed as
\begin{align}\label{minimize}\
{\left| {{{\bf{\beta}} _{a,w}}} \right|^2} + {{\bf{v}}^H}{{\bf{\Psi}} _{a,w}}{\beta _{a,w}} + {\beta _{a,w}}{{\bf{\Psi}} ^H}_{a,w}{\bf{v}} + {{\bf{v}}^H}{{\bf{\Psi}} _{a,w}}{{\bf{{{\Psi}}}} ^H}_{a,w}{\bf{v}}\nonumber \\
 \le N{\lambda _{\max }}\left( {{{\bf{\Psi}} _{a,w}}{{\bf{\Psi}} ^H}_{a,w}} \right) + 2{\mathop{\rm Re}\nolimits} \left( {{{\bf{v}}^H}{{\bf{\Psi}} _{a,w}}{\beta _{a,w}}} \right) + {\left| {{\beta _{a,w}}} \right|^2}
\end{align}
{where the equality holds when $\bf{v}$ is the eigenvector corresponding to the maximum eigenvalue ${\lambda _{\max }}\left( {{{\bf{\Psi}} _{a,w}}{{\bf{\Psi}} ^H}_{a,w}} \right)$}. To minimize the Willie's link power gain, we can minimize its upper bound. According to \eqref{minimize}, the optimal phase shifts vector is calculated by minimizing   ${\mathop{\rm Re}\nolimits} \left( {{{\bf{v}}^H}{{\bf{\Psi}} _{a,w}}{\beta _{a,w}}} \right) $, which is given by
 \begin{align}\label{optimaltheta}\
&{\mathop{\rm Re}\nolimits} \left( {{{\bf{v}}^H}{{\bf{\Psi}} _{a,w}}{\beta _{a,w}}} \right)= \sum\limits_{i = 1}^N {\left| {{g^H}_{{s_i},w}} \right|} \left| {{{\bf{h}}_{a,{s_i}}}{{{\bf{w}} }}} \right|\left| {{{\bf{h}}^H}_{a,w}{{{\bf{w}} }}} \right|\nonumber\\
&\quad\quad\quad\quad \times \cos \left( {{\theta _i} + \arg \left( {{g^H}_{{s_i},w}} \right) + \arg \left( {{{\bf{h}}_{a,{s_i}}}{{{\bf{w}} }}} \right) - \arg \left( {{{\bf{h}}^H}_{a,w}{{{\bf{w}} }}} \right)} \right).
  \end{align}
When $\cos \left( {{\theta _i} + \arg \left( {{g^H}_{{s_i},w}} \right) + \arg \left( {{{\bf{h}}_{a,{s_i}}}{{{\bf{w}} }}} \right) - \arg \left( {{{\bf{h}}^H}_{a,w}{{{\bf{w}} }}} \right)} \right)=-1$, the upper bound on Willie's link power gain is minimized. Thus, the optimal ${\theta _i}^*$ is given by
 \begin{align}\label{optimaltheta1}\
 {\theta _i}^* = \pi+\arg \left( {{{\bf{h}}^H}_{a,w}{\bf{w}} } \right) - \arg \left( {{g^H}_{{s_i},w}} \right) - \arg \left( {{{\bf{h}}}_{a,{s_i}}{{{\bf{w}} }}} \right).
  \end{align}
 Then, perform iterations between \eqref{optimalomega} and \eqref{optimaltheta1} until the accuracy of the covert transmission rate is satisfied. Note that in this suboptimal algorithm, we should first calculate the optimal beamforming vector for a given $\bf{v}$. Then, the optimal phase shifts vector is calculated for a given beamforming vector. This is because for a random $\bf{w}$, there might not exist a phase shifts vector $\bf{v}$ satisfying the covert constraint. In addition, Since the exact expressions of phase shifts and transmit beamforming vectors are given by \eqref{optimalomega} and \eqref{optimaltheta1}, respectively, the complexity for each iteration is  $O\left( {M + N} \right)$.
 \section{Covert transmission with imperfect CSI}
Since Willie can be unfriendly to Alice and deliberately hides himself, Alice cannot obtain the full instantaneous CSI of Willie's link perfectly. Hence, we investigate the covert transmission with imperfect CSI of Willie's link. In the following, we will investigate the covert transmission with imperfect CSI of the Alice-to-Wilie link, the IRS-to-Willie link, and the Alice-to-IRS link, respectively. { In addition, the impact of imperfect cascaded CSI of Willie's link on covert transmission is omitted due to space limitation \footnote{{If the estimated cascaded CSI of the Alice-IRS-Willie
and Alice-IRS-Bob links are independent, our proposed algorithm can handle the case with imperfect
cascaded CSI of Alice-IRS-Willie link; otherwise, it is challenging to obtain the maximum covert rate using
our proposed algorithm.}}.}
 \subsection{Imperfect CSI of the Alice-to-Willie Link}
Assume that the channels between Alice and Willie are subject to error,
 {denoted by ${{\Delta}}{\bf{h}}_{a,w}$, which is norm-bounded, i.e., $\|{{\Delta}}{\bf{h}}_{a,w}\|\leq \zeta_{a,w}$. Then, the estimated Alice-to-Willie link is expressed as ${\tilde{\bf{h}}}_{a,w}{=}{\bf{h}}_{a,w}+{{\Delta}}{\bf{h}}_{a,w}$}. In this case, the signal power received at Willie should be less than $\eta$, and the covert constraint is rewritten as
  \begin{align}\label{estimate1}\
  {{{\left| {\left( {{{\bf{h}}^H}_{a,w} +{{\Delta}}{{\bf{h}}^H}_{a,w}+ {{\bf{v}}^H} diag \left( {{{\bf{g}}^H_{s,w}}} \right){\bf{h}}_{a,s}} \right)\bf{{\bf{w}}} } \right|}^2} \le \eta }.
   \end{align}
The following inequalities hold
\begin{align}\label{awestimate2}\
&\left| {\left( {{{\bf{h}}^H}_{a,w} + \Delta {{\bf{h}}^H}_{a,w} + {{\bf{v}}^H} diag \left( {{{\bf{g}}^H_{s,w}}} \right){\bf{h}}_{a,s}} \right){\bf{w}}} \right| \nonumber \\
 &\quad \overset{(a)}\le \left| {\left( {{{\bf{h}}^H}_{a,w} + {{\bf{v}}^H} diag \left( {{{\bf{g}}^H_{s,w}}} \right){\bf{h}}_{a,s}} \right){\bf{w}}} \right|{{ + }}\left| {\Delta {{\bf{h}}^H}_{a,w}{\bf{w}}} \right|\nonumber  \\
 &\quad \overset{(b)} \le \left| {\left( {{{\bf{h}}^H}_{a,w} + {{\bf{v}}^H} diag \left( {{{\bf{g}}^H_{s,w}}} \right){\bf{h}}_{a,s}} \right){\bf{w}}} \right|{{ + }}\left\| {\Delta {{\bf{h}}^H}_{a,w}} \right\|\left\| {\bf{w}} \right\|\nonumber  \\
&\quad {\leq} \left| {\left( {{{\bf{h}}^H}_{a,w} + {{\bf{v}}^H} diag \left( {{{\bf{g}}^H_{s,w}}} \right){\bf{h}}_{a,s}} \right){\bf{w}}} \right|{\rm{ + }}\zeta_{a,w} \left\| {\bf{w}} \right\|
   \end{align}
where (a) is obtained by using the triangle inequality and (b) is obtained by applying the Cauchy-Schwarz inequality \cite{KTPhan}. Specifically, the equality of (a) holds when \\
$\arg \left( {\left( {{\bf{h}}_{a,w}^H + {{\bf{v}}^H}diag\left( {{\bf{g}}_{s,w}^H} \right){{\bf{h}}_{a,s}}} \right){\bf{w}}} \right) = \arg \left( {\Delta {\bf{h}}_{a,w}^H {\bf{w}}} \right)$, and the equality of (b) holds when ${\Delta {\bf{h}}_{a,w}^H}$ and $\bf{w}$ are linear correlation.  Then, according to \eqref{awestimate2}, the signal power received at Willie satisfies the following inequalities
\begin{align}\label{awestimate3}\
{\left| {\left( {{{\bf{h}}^H}_{a,w} + \Delta {{\bf{h}}^H}_{a,w} + {{\bf{v}}^H} diag \left( {{{\bf{g}}^H_{s,w}}} \right){\bf{h}}_{a,s}} \right){\bf{w}}} \right|^{\rm{2}}}\leq {{\bf{w}}^H}{{{\bf{\tilde U}}}_{a,w}}{\bf{w}}
   \end{align}
where ${{{\bf{\tilde U}}}_{a,w}} ={{{\bf{ U}}}_{a,w}} {{\rm{ + }}\left( {{\zeta_{a,w} ^{\rm{2}}}{\rm{ + }}2\zeta_{a,w}  \left\| {\left( {{{\bf{h}}^H}_{a,w} + {{\bf{v}}^H} diag \left( {{{\bf{g}}^H_{s,w}}} \right){\bf{h}}_{a,s}} \right)} \right\|} \right){\bf{I}}_M}$,\\
 and ${{{\bf{ U}}}_{a,w}} ={{\left( {{{\bf{h}}^H}_{a,w} + {{\bf{v}}^H} diag \left( {{{\bf{g}}^H_{s,w}}} \right){\bf{h}}_{a,s}} \right)}^H}\left( {{{\bf{h}}^H}_{a,w} + {{\bf{v}}^H} diag \left( {{{\bf{g}}^H_{s,w}}} \right){\bf{h}}_{a,s}} \right)$.
Since the phase shifts vector  $\bf{v}$ and transmit beamforming vector $\bf{w}$ are coupled, we still use the alternating algorithm to solve the optimization problem. First, for a given phase shifts vector $\bf{v}$, let ${\bf{W}} = {\bf{w}}{{\bf{w}}^H}$, then, the optimization problem is formulated as
   \begin{align}\label{imaw1}\
&\mathop {\max }\limits_{{\bf{w}} }\quad Tr\left( {{{\bf{U}}_{a,b}}{\bf{W}}} \right) \nonumber \\
& s.t. \quad {Tr\left( {{\bf{W}} } \right) \le {P_{\max }}} \nonumber \\
&\quad\quad Tr\left( {{{\bf{\tilde{U}}}_{a,w}}{\bf{W}}} \right) \le \eta, {\bf{W}} \succeq 0, rank\left( {{\bf{W}} } \right)=1
 \end{align}
where ${{\bf{U}}_{a,b}} = {\left( {{{\bf{h}}^H}_{a,b} + {{\bf{v}}^H} diag \left( {{{\bf{g}}^H_{s,b}}} \right){\bf{h}}_{a,s}} \right)^H}\left( {{{\bf{h}}^H}_{a,b} + {{\bf{v}}^H} diag \left( {{{\bf{g}}^H_{s,w}}} \right){\bf{h}}_{a,s}} \right)$. When neglecting the constraint $rank\left( {{\bf{W}} } \right)=1$, SDP can be used to solve the optimization problem. Moreover, Gaussian randomization is used to obtain the optimal $\bf{w}$ satisfying the rank 1 constraint. After the beamforming vector $\bf{w}$ is obtained, we derive next the phase shifts vector with a fixed $\bf{w}$. For a given  beamforming vector $\bf{w}$,
the optimization problem is reformulated as
\begin{align}\label{imaw2}\
&{\mathop {\max }\limits_{\bf{v}} }\quad {{{\left| {{\alpha _{a,b}}} \right|}^2} + Tr\left( {{{\bf{R}}_{a,b}}{\bf{V}}} \right)}
\nonumber \\
&s.t.\quad  {{{\left| {{\alpha _{a,w}}} \right|}^2} + Tr\left( {{{\bf{R}}_{a,w}}{\bf{V}}} \right) \le {\left( {\sqrt \eta   - \zeta_{a,w} \left\| {\bf{w}} \right\|} \right)^2} } \nonumber\\
&\quad \quad {{{\bf{V}}_{n,n}} = 1},{{\bf{V}} \succeq 0},  rank\left( {\bf{V}}\right)=1
 \end{align}
where the covert constraint is obtained using \eqref{awestimate2}. Similar to \eqref{perproblem3}, the optimal phase shifts vector can be obtained  using the SDP and Gaussian randomization techniques. The complexity of \eqref{imaw1} is $O\left( {{{\left( {{M^{\rm{2}}}{\rm{ + 1}}} \right)}^{{\rm{3}}{\rm{.5}}}}} \right)$ for each iteration, and  the complexity of \eqref{imaw2} is  $O\left( {{{\left( {{{\left( {N + 1} \right)}^{\rm{2}}}{\rm{ + 1}}} \right)}^{{{3}}{{.5}}}}} \right)$ for each iteration. Then, we iterate between \eqref{imaw1} and \eqref{imaw2} until the gap of covert transmission rate between the \emph{i}-th and $(i+1)$-th iteration is less than a predetermined threshold $\Gamma$. The detail of the algorithm is presented in the following table. In the Algorithm 1, we should first derive the beamforming vector with a fixed phase shift. If we first derive the phase shifts with a fixed beamforming vector, the covert constraint cannot be satisfied.

\begin{algorithm}[h]
\caption{Maximize the Covert Transmit Rate With imperfect CSI of Alice-to-Willie link }
\begin{algorithmic}[1]
\STATE {Initialization: $N$, $M$, $\eta$, $P_{\max}$}.
\STATE {Randomly set the phase shift $\bf{v}$, let $f_0{(\bf{w}, \bf{v})}$=0, and set $i=1$}.
\STATE {For a given $\bf{v}$, calculate the optimal $\bf{W}$ according to \eqref{imaw1}}.
\STATE {Perform Gaussian randomization for $\bf{W}$, and obtain the optimal transmit beamforming vector ${\bf{w}}^*$}.
\STATE {For the given ${\bf{w}}^*$, calculate the optimal $\bf{V}$ according to \eqref{imaw2}.}
\STATE {Perform Gaussian randomization for $\bf{V}$, and obtain the optimal phase shifts vector ${\bf{v}}^*$.}
\STATE  {Calculate the covert transmit rate $f_i{(\bf{w}^*, \bf{v}^*)}$}.
\STATE {If $f_i{(\bf{w}^*, \bf{v}^*)}-f_{i-1}{(\bf{w}, \bf{v})}\leq \Gamma $, break;
else $i=i+1$, update $\bf{w}=\bf{w}^*$ and $\bf{v}=\bf{v}^*$,  and repeat step 3 until the accuracy of the covert rate is satisfied}.
\end{algorithmic}
\end{algorithm}
\subsection{Imperfect CSI of the IRS-to-Willie Link}
Assume that the estimated CSI of the IRS-to-Willie links is imperfect, and subject to error,
denoted by ${{\Delta}}{\bf{g}}_{s,w}$, which is norm-bounded, i.e., $\|{{\Delta}}{\bf{g}}_{s,w}\|\leq \zeta_{s,w}$. The estimated IRS-to-Willie link is ${{\bf{\tilde g}}^H}_{s,w} ={{{\bf{g}}^H}_{s,w} + \Delta {{\bf{g}}^H}_{s,w}}$. Similar to \eqref{awestimate2} and \eqref{awestimate3}, the signal power received at Willie satisfies the following inequality
\begin{align}\label{swestimate}\
{\left| {\left( {{{\bf{h}}^H}_{a,w} + \left( {{{\bf{g}}^H}_{s,w} + \Delta {{\bf{g}}^H}_{s,w}} \right){\bf{\Theta }}{{\bf{h}}_{a,s}}} \right){\bf{w}}} \right|^2}\le {{\bf{w}}^H}{{{\bf{\hat U}}}_{a,w}}{\bf{w}}
\end{align}
where ${{{\bf{\hat U}}}_{a,w}} = {{{\bf{ U}}}_{a,w}}{ + \left( {{\zeta^2}_{s,w}{{\left\| {{\bf{\Theta }}{{\bf{h}}_{a,s}}} \right\|}^2} + 2{\zeta_{s,w}}\left\| {{\bf{\Theta }}{{\bf{h}}_{a,s}}} \right\|\left\| {\left( {{{\bf{h}}^H}_{a,w} + {{\bf{g}}^H}_{s,w}{\bf{\Theta }}{{\bf{h}}_{a,s}}} \right)} \right\|} \right){\bf{I}}_M} $. Then, for a given phase shifts vector $\bf{v}$, the optimization problem is given by
  \begin{align}\label{imswproblem}\
&\mathop {\max }\limits_{{\bf{w}} } \quad Tr\left( {{{\bf{{U}}}_{a,b}}{\bf{W}}} \right) \nonumber \\
& s.t. \quad {Tr\left( {{\bf{W}} } \right) \le {P_{\max }}}\nonumber \\
&\quad\quad Tr\left( {{{\bf{\hat{U}}}_{a,w}}{\bf{W}}} \right) \le \eta, {\bf{W}} \succeq 0, rank\left( {{\bf{W}} } \right)=1.
 \end{align}
For \eqref{imswproblem}, neglect the rank-1 constraint, and SDP can be used to obtain the optimal $\bf{W}$. Then, Gaussian randomization is applied to obtain the optimal beamforming vector $\bf{w}$. For the calculated $\bf{w}$, we optimize the phase shifts for covert transmission rate maximization. Note that for the imperfect ${{\bf{g}}^H}_{s,w}$, the following inequalities are satisfied
\begin{align}\label{imswv1}\
&\left| {\left( {{{\bf{h}}^H}_{a,w} + \left( {{{\bf{g}}^H}_{s,w} + \Delta {{\bf{g}}^H}_{s,w}} \right){\bf{\Theta }}{{\bf{h}}_{a,s}}} \right){\bf{w}}} \right|\overset{(a)}\le \left| {\left( {{{\bf{h}}^H}_{a,w} + {{\bf{g}}^H}_{s,w}{\bf{\Theta }}{{\bf{h}}_{a,s}}} \right){\bf{w}}} \right| + {\zeta _{s,w}}\left\| {{\bf{\Theta }}{{\bf{h}}_{a,s}}{\bf{w}}} \right\|\nonumber \\
 & \quad\quad\quad\quad\quad\quad\quad\quad\quad\quad\quad\overset{(b)} \le \left| {{{\bf{h}}^H}_{a,w}{\bf{w}} + {{\bf{v}}^{\rm{H}}}{\rm{diag}}\left( {{{\bf{g}}^H}_{s,w}} \right){{\bf{h}}_{a,s}}{\bf{w}}} \right| + {\zeta _{s,w}}\left\| {{{\bf{h}}_{a,s}}{\bf{w}}} \right\|
\end{align}
where (a) is obtained by applying the triangle and the Cauchy-Schwarz inequalities; (b) holds due to $\left\| {{\bf{\Theta }}{{\bf{h}}_{a,s}}{\bf{w}}} \right\|{{ = }}\left\| {{{\bf{h}}_{a,s}}{\bf{w}}} \right\|$. Then, according to \eqref{imswv1}, the received signal power at Willie is given by
\begin{align}\label{problem4}\
&{\left| {\left( {{{\bf{h}}^H}_{a,w} + \left( {{{\bf{g}}^H}_{s,w} + \Delta {{\bf{g}}^H}_{s,w}} \right){\bf{\Theta }}{{\bf{h}}_{a,s}}} \right){\bf{w}}} \right|^2}\leq Tr\left( { {{{\bf{R}}_{a,w}} }{\bf{V}}} \right) + \Lambda_{s,w}
\end{align}
 where $ \Lambda_{s,w}={\left| {{{\bf{h}}^H}_{a,w}{\bf{w}}} \right|^2} + {\zeta ^{\rm{2}}}_{s,w}{\left\| { {{{\bf{h}}_{a,s}}{\bf{w}}} } \right\|^{\rm{2}}}{\rm{ + 2}}{\zeta _{s,w}} \left\| {{{{\bf{h}}_{a,s}}{\bf{w}}}} \right\|\left( {\left| {{{\bf{h}}^H}_{a,w}{\bf{w}}} \right|{\rm{ + }}\| {{{\bf{g}}^H}_{s,w}}\|\left\| {{\bf{h}}_{a,s}}{\bf{w}} \right\|} \right)$. Therefore, for a given $\bf{w}$, the optimization problem is expressed as
\begin{align}\label{imswproblemv}\
&{\mathop {\max }\limits_{\bf{v}} }\quad { Tr\left( {{{\bf{R}}_{a,b}}{\bf{V}}} \right)}\nonumber \\
& {s.t.} \quad Tr\left( { {{{\bf{R}}_{a,w}}} {\bf{V}}} \right) \leq \eta - \Lambda_{s,w} \nonumber\\
&\quad \quad  {{{\bf{V}}_{n,n}} = 1},{{\bf{V}} \succeq 0},rank\left(\bf{V}\right)=1.
 \end{align}
 Similarly, SDP and Gaussian randomization are used to solve the problem. Eqs. \eqref{imswproblem} and \eqref{imswproblemv} are iteratively solved  until the gap of covert transmission rate between the \emph{i}-th and the (\emph{i}+1)-th iteration is less than a predetermined threshold $\Gamma$.
 \subsection{Imperfect CSI of the Alice-to-IRS Link}
 When the Alice-to-IRS link is not estimated perfectly, both Bob's and Willie's links are affected. Assume that the estimated CSI ${{\bf{\tilde h}}}^H_{a,s} ={{{\bf{h}}^H}_{a,s} + \Delta {{\bf{h}}^H}_{a,s}}$, where the estimation error $\Delta {{\bf{h}}^H}_{a,s}$ is norm-bounded as $\|\Delta {{\bf{h}}^H}_{a,s}\|\leq \zeta_{a,s}$. For a given phase shifts vector $\bf{v}$, the received signal power at Willie and Bob are, respectively, expressed as
\begin{align}\label{IIRS1}\
{\left| {\left( {{{\bf{h}}^H}_{a,b} + {{\bf{g}}^H}_{s,b}{\bf{\Theta }}\left( {{{\bf{h}}_{a,s}} + \Delta {{\bf{h}}_{a,s}}} \right)} \right){\bf{w}}} \right|^2}&\ge {\left( {\left| {\left( {{{\bf{h}}^H}_{a,b} + {{\bf{g}}^H}_{s,b}{\bf{\Theta }}{{\bf{h}}_{a,s}}} \right){\bf{w}}} \right| - \left| {{{\bf{g}}^H}_{s,b}{\bf{\Theta }}\Delta {{\bf{h}}_{a,s}}{\bf{w}}} \right|} \right)^2}\nonumber\\
&\ge {{\bf{w}}^H}{{{\bf{\bar U}}}_{a,b}}{\bf{w}}
 \end{align}
and
\begin{align}\label{IIRS2}\
{\left| {\left( {{{\bf{h}}^H}_{a,w} + {{\bf{g}}^H}_{s,w}{\bf{\Theta }}\left( {{{\bf{h}}_{a,s}} + \Delta {{\bf{h}}_{a,s}}} \right)} \right){\bf{w}}} \right|^2} &\leq {\left| {\left( {{{\bf{h}}^H}_{a,w} + {{\bf{g}}^H}_{s,w}{\bf{\Theta }}{{\bf{h}}_{a,s}}} \right){\bf{w}}} \right|^2} + {\zeta ^2}_{a,s}{\left\| {{{\bf{g}}^H}_{s,w}{\bf{\Theta }}} \right\|^2}{\left\| {\bf{w}} \right\|^2}\nonumber \\
 &\quad + 2{\zeta_{a,s}}\left\| {{{\bf{g}}^H}_{s,w}{\bf{\Theta }}} \right\|\left\| {\left( {{{\bf{h}}^H}_{a,w} + {{\bf{g}}^H}_{s,w}{\bf{\Theta }}{{\bf{h}}_{a,s}}} \right)} \right\|{\left\| {\bf{w}} \right\|^2}\nonumber \\
 &\leq {{\bf{w}}^H}{{{\bf{\bar{U}}}}_{a,w}}{\bf{w}}
 \end{align}
 where ${{{\bf{\bar{U}}}}_{a,w}} = {{{\bf{{U}}}}_{a,w}}+ \left( {{\zeta^2}_{a,s}{{\left\| {{{\bf{g}}^H}_{s,w}{\bf{\Theta }}} \right\|}^2} + 2{\zeta_{a,s}}\left\| {{{\bf{g}}^H}_{s,w}{\bf{\Theta }}} \right\|\left\| {\left( {{{\bf{h}}^H}_{a,w} + {{\bf{g}}^H}_{s,w}{\bf{\Theta }}{{\bf{h}}_{a,s}}} \right)} \right\|} \right){\bf{I}}_M$, and
${{{\bf{\bar U}}}_{a,b}} = {{{\bf{ U}}}_{a,b}}+ \left( {{\zeta^2}_{a,s}{{\left\| {{{\bf{g}}^H}_{s,b}{\bf{\Theta }}} \right\|}^2} - 2{\zeta_{a,s}}\left\| {{{\bf{g}}^H}_{s,b}{\bf{\Theta }}} \right\|\left\| {\left( {{{\bf{h}}^H}_{a,b} + {{\bf{g}}^H}_{s,b}{\bf{\Theta }}{{\bf{h}}_{a,s}}} \right)} \right\|} \right){\bf{I}}_M$. Thus, according to the lower bound on the signal power received at Bob, \eqref{IIRS1} and the upper bound on the signal power received at Willie given in \eqref{IIRS2}, for a given phase shift vector, the optimization problem is given by
\begin{align}\label{IIRS3f}\
&\mathop {\max }\limits_{{\bf{W}} } Tr\left( {{{\bf{\bar{U}}}_{a,b}}{\bf{W}}} \right) \nonumber \\
&s.t.\quad {Tr\left( {{\bf{W}} } \right) \le {P_{\max }}} \nonumber \\
&\quad \quad  Tr\left( {{{\bf{\bar{U}}}_{a,w}}{\bf{W}}} \right) \le \eta, {\bf{W}} \succeq 0, rank\left( {{\bf{W}} } \right)=1.
\end{align}
SDP and Gaussian randomization are performed to calculate the optimal $\bf{w}$. For the calculated $\bf{w}$, the received signal power at Bob and Willie satisfy the following inequalities
\begin{align}\label{IIRS3}\
&{\left| {\left( {{{\bf{h}}^H}_{a,b} + {{\bf{g}}^H}_{s,b}{\bf{\Theta }}\left( {{{\bf{h}}_{a,s}} + \Delta {{\bf{h}}_{a,s}}} \right)} \right){\bf{w}}} \right|^2}\ge {{\bf{\bar{v}}}^H}{{\bf{R}}_{a,b}}{\bf{\bar{v}}}{\rm{ + }}{\zeta ^2}_{a,s}{\left\| {\bf{w}} \right\|^2}{\left\| {diag\left( {{{\bf{g}}^H}_{s,b}} \right)} \right\|^{\rm{2}}}{\rm{ + }}{\left| {{{\bf{h}}^H}_{a,b}{\bf{w}}} \right|^{\rm{2}}}\nonumber\\
 &\quad\quad\quad\quad\quad\quad - 2{\zeta _{a,s}} \left\| { {{{\bf{g}}^H}_{s,b}} } \right\|\left\| {\bf{w}} \right\|\left( {\left| {{{\bf{h}}^H}_{a,b}{\bf{w}}} \right| + \| {{{\bf{g}}^H}_{s,b}} \|\left\| {{{\bf{h}}_{a,s}}{\bf{w}}} \right\|} \right)
\end{align}
and
\begin{align}\label{IIRS4}\
{\left| {\left( {{{\bf{h}}^H}_{a,w} + {{\bf{g}}^H}_{s,w}{\bf{\Theta }}\left( {{{\bf{h}}_{a,s}} + \Delta {{\bf{h}}_{a,s}}} \right)} \right){\bf{w}}} \right|^2} \le {{\bf{\bar{v}}}^H}{{\bf{R}}_{a,w}}{\bf{\bar{v}}} + {\Lambda _{a,s}}
\end{align}
where ${\Lambda _{a,s}} = {\zeta ^2}_{a,s}{\left\| {\bf{w}} \right\|^2}{\left\| {{{\bf{g}}^H}_{s,w}} \right\|^{\rm{2}}}{\rm{ + }}{\left| {{{\bf{h}}^H}_{a,w}{\bf{w}}} \right|^{\rm{2}}}\nonumber \\
+ 2{\zeta _{a,s}} \left\| {{{\bf{g}}^H}_{s,w}} \right\|\left\| {\bf{w}} \right\|\left( {\left| {{{\bf{h}}^H}_{a,w}{\bf{w}}} \right| + \| {{{\bf{g}}^H}_{s,w}} \| \left\| {{{\bf{h}}_{a,s}}{\bf{w}}} \right\|} \right)$. Then, for a beamforming vector $\bf{w}$, the optimization problem is rewritten as
\begin{align}\label{problem3}\
&{\mathop {\max }\limits_{\bf{V}} }\quad { Tr\left( {{{\bf{R}}_{a,b}}{\bf{V}}} \right)}\nonumber \\
&s.t.\quad Tr\left( {{{{\bf{R}}_{a,w}}} {\bf{V}}} \right) \leq \eta - \Lambda_{a,s} \nonumber\\
&\quad\quad {{{\bf{V}}_{n,n}} = 1},{{\bf{V}} \succeq 0}, rank \left( {\bf{V}}\right) =1.
 \end{align}
 SDR and Gausssian randomization are used to solve \eqref{problem3}. Finally,  Eqs. \eqref{IIRS3f} and \eqref{problem3} are iteratively solved until the gap of covert transmission rate between the \emph{i}-th and (\emph{i}+1)-th iteration is less than a predetermined threshold $\Gamma$.
  \subsection{Imperfect CSI of Both the IRS-to-Willie and Alice-to-Willie Links}
 { When both the Alice-to-Willie and IRS-to-Willie links are imperfect, i.e.,  $\|{{\Delta}}{\bf{h}}_{a,w}\|\leq \zeta_{a,w}$, and $\|{{\Delta}}{\bf{g}}_{s,w}\|\leq \zeta_{s,w}$, the maximum covert transmission rate can also be derived iteratively. For a given phase shifts vector $\bf{v}$, the optimization problem is formulated as
\begin{align}\label{cimaw1}\
&\mathop {\max }\limits_{{\bf{w}} }\quad Tr\left( {{{\bf{U}}_{a,b}}{\bf{W}}} \right) \nonumber \\
& s.t. \quad {Tr\left( {{\bf{W}} } \right) \le {P_{\max }}} \nonumber \\
&\quad\quad Tr\left( {{{\bf{\breve{U}}}_{a,w}}{\bf{W}}} \right) \le \eta, {\bf{W}} \succeq 0, rank\left( {{\bf{W}} } \right)=1
\end{align}
where $ {{\bf{\breve{U}}}_{a,w}}= \left( {2\left( {{\zeta _{s,w}}\left\| {{{\bf{h}}_{a,s}}} \right\| + {\zeta _{a,w}}} \right)\left\| {\left( {{{\bf{h}}^H}_{a,w} + {{\bf{v}}^H}diag({{\bf{g}}^H}_{s,w}){{\bf{h}}_{a,s}}} \right)} \right\| + {{\left( {{\zeta _{s,w}}\left\| {{{\bf{h}}_{a,s}}} \right\|+ {\zeta _{a,w}}} \right)}^2}} \right){{\bf{I}}_M}\nonumber \\
+ {{\bf{U}}_{a,w}} $. Then, problem in \eqref{cimaw1} can be solved by SDR and Gaussian randomization. For a given beamforming vector $\bf{w}$, the optimization problem is rewritten as
\begin{align}\label{cimswproblemv}\
&{\mathop {\max }\limits_{\bf{v}} }\quad { Tr\left( {{{\bf{R}}_{a,b}}{\bf{V}}} \right)}\nonumber \\
& {s.t.} \quad Tr\left( { {{{\bf{R}}_{a,w}}} {\bf{V}}} \right) \leq \eta - \Lambda_{a,w} \nonumber\\
&\quad \quad  {{{\bf{V}}_{n,n}} = 1},{{\bf{V}} \succeq 0},rank\left(\bf{V}\right)=1.
 \end{align}
where ${\Lambda _{a,w}} = {\left| {{{\bf{h}}^H}_{a,w}{\bf{w}} } \right|^{{2}}}{{ + }}{\left( {{\zeta _{s,w}}\left\| {{{\bf{h}}_{a,s}}{\bf{w}} } \right\| + {\zeta _{a,w}}\left\| \bf{w}  \right\|} \right)^{\rm{2}}}\nonumber\\
{\rm{ + 2}}\left( {{\zeta _{s,w}}\left\| {{{\bf{h}}_{a,s}}{\bf{w}} } \right\| + {\zeta _{a,w}}\left\| {\bf{w}} \right\|} \right)\left( {\left| {{{\bf{h}}^H}_{a,w}{\bf{w}} } \right| + \left\| {{{\bf{g}}^H}_{s,w}} \right\|\left\| {{{\bf{h}}_{a,s}}{\bf{w}} } \right\|} \right)$.
Similarly, problem \eqref{cimswproblemv} is solved by SDR and Gaussian randomization. Eqs. \eqref{cimaw1} and \eqref{cimswproblemv} are iteratively solved until the gap of the covert transmission rate between the \emph{i}-th and (\emph{i}+1)-th iterations is less than a predetermined threshold $\Gamma$.}

\section{Numerical Results}
\begin{figure}
\centering
\includegraphics[width=0.4\textwidth]{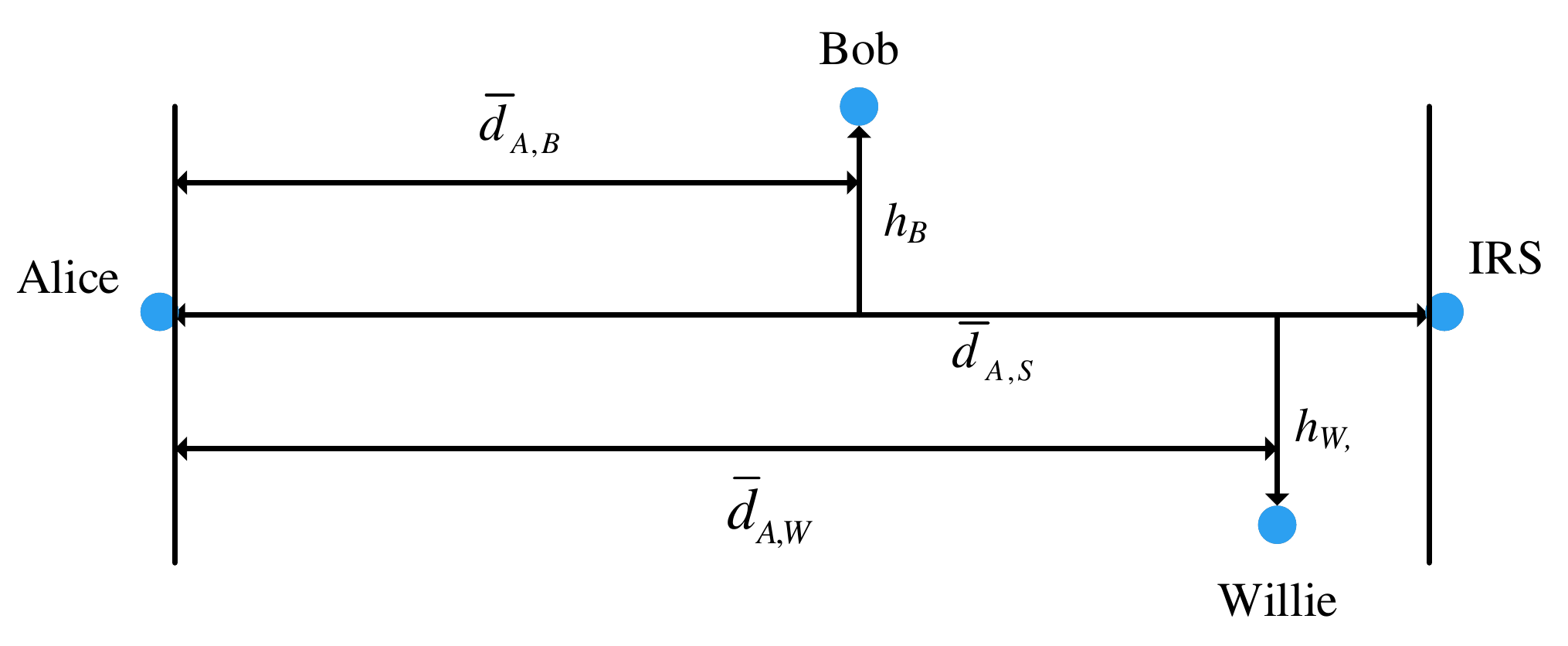}
\caption{The simulation model.} \label{Fig2}
\end{figure}

In this section, numerical results for covert transmission with IRS assistance are presented. Moreover, covert transmission without IRS is simulated as a benchmark. We set ${P_{max}}=10$ dBm,  ${{{{\tilde \sigma }^2}_w}}={{{{\sigma }^2}_b}}=-90$ dBm, and $\Gamma=10^{-4}$. The small-scale fading of all  channels follows the Rayleigh fading model. The path loss model is $PL = \left( {P{L_0} - 10{{\log }_{10}}{{\left( {\frac{d}{{{d_0}}}} \right)}^\mu }} \right)$ dBm, where $P{L_0}=-30$ $\mathbf{dB}$, is the path loss with reference distance  ${d_0}=1$ m, $\mu$ is the path loss exponent, and $d$ is the distance between the transmitter and receiver. {The average iteration number for the optimal algorithm is about 30, and the number of Gaussian randomizations is 1000}. The simulation model is presented in Fig. \ref{Fig2}, where Alice and the IRS lie along the same line, and the horizontal distance between Alice and the IRS is ${\bar{d}_{A,S}}$. The horizontal and vertical distances between Alice and Willie are denoted as ${{\bar{d}}_{A,W}}$ and $h_{w}$, respectively. The distance between Alice and Willie is  ${{{d}}_{A,W}}=\sqrt{\left({{\bar{d}}^2_{A,W}}+h^2_{w}\right)}$.  The horizontal and vertical distances between Alice and Bob are denoted as $\bar{d}_{A,B}$ and ${{{h}}_{B}}$, respectively. The distance between Alice and Bob is  ${{{d}}_{A,B}}=\sqrt{\left({{\bar{d}}^2_{A,B}}+h^2_{B}\right)}$.  The path loss exponents of the Alice-to-IRS, the Alice-to-Bob, the Alice-to-Willie, the IRS-to-Bob and the IRS-to-Willie links are $\mu_{AI}$, $\mu_{AB}$, $\mu_{AW}$, $\mu_{IB}$, and $\mu_{IW}$, respectively.
\begin{figure}
\centering
\includegraphics[width=0.4\textwidth]{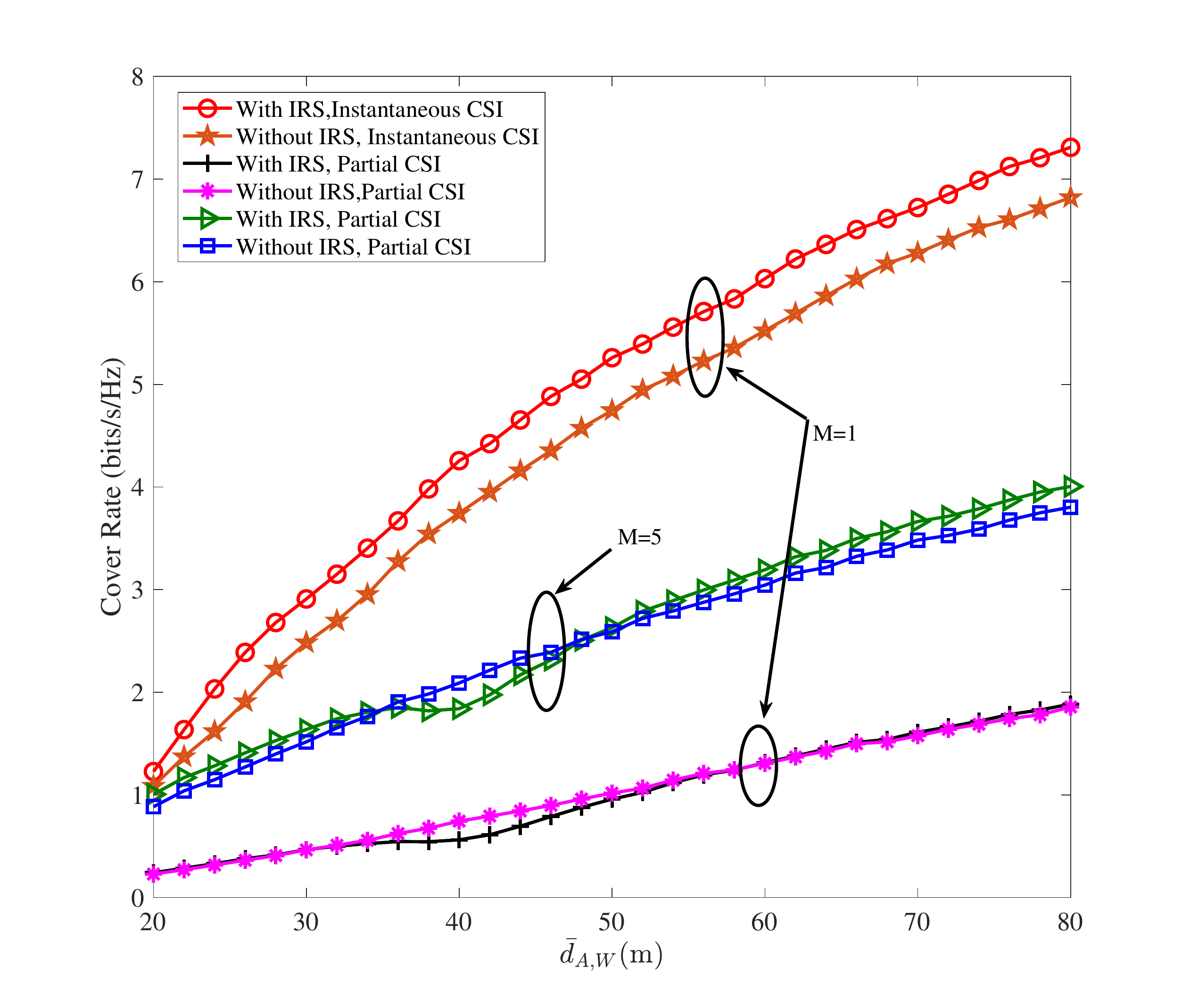}
\caption{The covert transmission rate against ${\bar{d}}_{A,W}$ with partial and instantaneous CSI of Willie's link, { $\rho=3$ $\mathbf{dB}$}, $\kappa=10^{-2}$,  $N=10$,  $h_W=5$ m, $h_B=3$ m, $\mu_{A,B}=2.5$, $\mu_{A,S}=\mu_{S,B}=2$, $\mu_{A,W}=\mu_{S,W}=2.5$,  and ${\bar{d}}_{A,B}={d}_{A,S}=40$ m.} \label{Fig3}
\end{figure}

We investigate the covert transmission with and without IRS assistance in Fig. \ref{Fig3}. When only the partial CSI of Willie's link available, we find that the covert transmission with IRS is not necessarily better than that without IRS whether single or multiple antennas are deployed at Alice. When Willie is close to the IRS, the allowed transmission power is constrained significantly and less than that without IRS. Although Bob's link power gain is strengthened with IRS assistance, Willie's link power gain is also strengthened even if the phase shifts vector is random for Willie's link in this case. When Willie is far away from the IRS, the covert transmission rate with IRS outperforms that without IRS. In this case, the impact of the IRS on Bob's link power gain is larger than that on Willie's link power gain. For comparison, the covert rate with instantaneous CSI of Willie's link is also illustrated in Fig. \ref{Fig3}. We can see that even if $M$=1, the covert rate with instantaneous CSI of Willie's link is much greater than that with partial CSI of Willie's link, which is consistent with our intuition. Hence, obtaining the instantaneous CSI of Willie's link is an important and effective way to improve the covert rate in practice. Moreover, when instantaneous CSI of all links are available, and the channel quality of Bob's link is comparable to that of Willie's link, the covert rate with IRS assistance is greater than that without IRS assistance.

\begin{figure}
\centering
\includegraphics[width=0.4\textwidth]{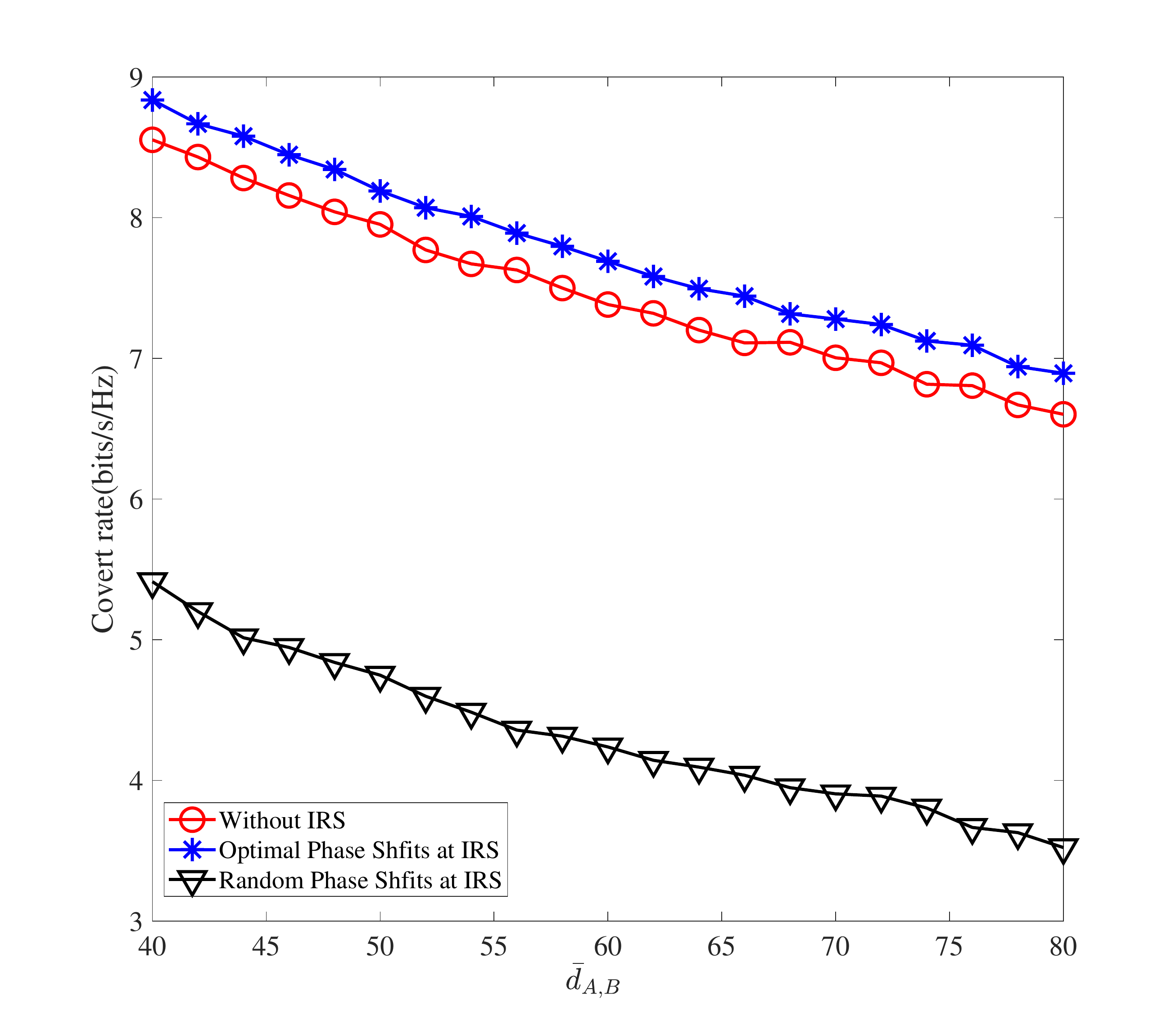}
\caption{The covert transmission rate against ${\bar{d}}_{A,B}$ with instantaneous CSI of Willie's link, $N=4$, $M=1$, $\rho=3$ $\mathbf{dB}$, $h_B=10$ m, $h_W=3$ m, $\mu_{A,B}=2$, $\mu_{A,W}=4.5$, $\mu_{S,B}=4.5$, $\mu_{S,W}=1.5$, $\mu_{A,S}=2$, and ${{d}}_{A,S}={d}_{A,W}=60$ m. } \label{Fig4}
\end{figure}

 For the case of a single antenna at Alice and full instantaneous CSI of  Willie's link available, Fig. \ref{Fig4} shows the covert transmission rate with and without IRS assistance. We find that the covert rate without IRS is greater than that with IRS. It is because the channel quality of the IRS-to-Bob link is much better than that of the IRS-to-Willie link, i.e.,  $\mu_{S,B}=4.5$ and $\mu_{S,W}=1.5$. In this case, although the signal power received at Bob is strengthened by the IRS, the signal power received at Willie is also strengthened by the IRS, which leads to the transmission power decrease at Alice. Moreover, under the case of IRS assistance, our proposed algorithm outperforms significantly the random phase shifts at IRS in terms of covert rate. This demonstrates the advantage of the proposed optimal algorithm.

\begin{figure}
\centering
\includegraphics[width=0.4\textwidth]{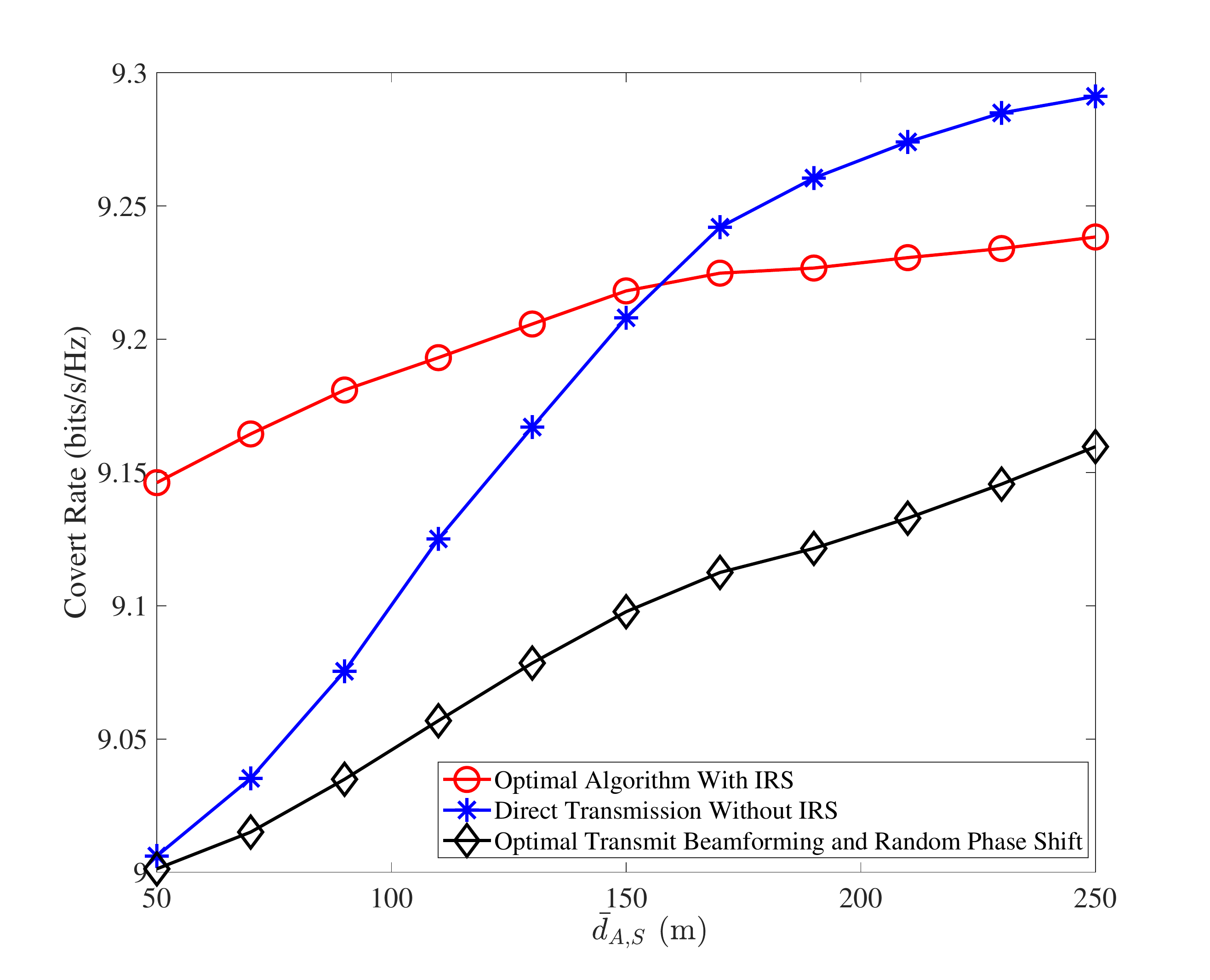}
\caption{The covert transmission rate against ${\bar{d}}_{A,S}$ with instantaneous CSI of Willie's link, ${\bar{d}}_{A,S}={\bar{d}}_{A,W}$, $\kappa=0.01$, $\mu_{A,B}=\mu_{A,S}=\mu_{S,W}=2$, $\mu_{A,W}=\mu_{S,B}=4$, $\rho=3$ $\mathbf{dB}$, ${\bar{d}}_{A,B}=200$ m, $M=5$, $N=10$, $h_{B}=200$ m, and $h_{W}=5$ m.} \label{Fig5}
\end{figure}

The covert rate against ${\bar{d}}_{A,S}$ is plotted when transmit beamforming is applied at Alice in Fig. \ref{Fig5}, where the distance between the IRS and Willie is fixed with $h_{W}$. The distance between the IRS and Bob is $\sqrt{\left(\left({{\bar{d}}_{A,B}}+{{\bar{d}}_{A,S}}\right))^2+h^2_{B}\right)}$, where Bob lies on the left side of Alice. We can see that the covert rate with IRS is greater than that without IRS when ${\bar{d}}_{A,S}$ is small, but the covert rate with IRS is less than that without IRS when ${\bar{d}}_{A,S}$ is large. When the IRS is close to Alice and Bob, the effect of IRS on Bob's link dominates the covert performance. However, when the IRS is far away from Alice and Bob, the effect of IRS on Willie's link dominates the covert transmission.  This demonstrates that the IRS is not necessarily beneficial to the covert transmission rate even if the transmit beamforming and phase shift are jointly optimized. In addition, we can see that our proposed optimal algorithm is better than the optimal transmit beamforming and random phase shift algorithm, which demonstrate the advantage of our proposed algorithm.

\begin{figure}
\centering
\includegraphics[width=0.4\textwidth]{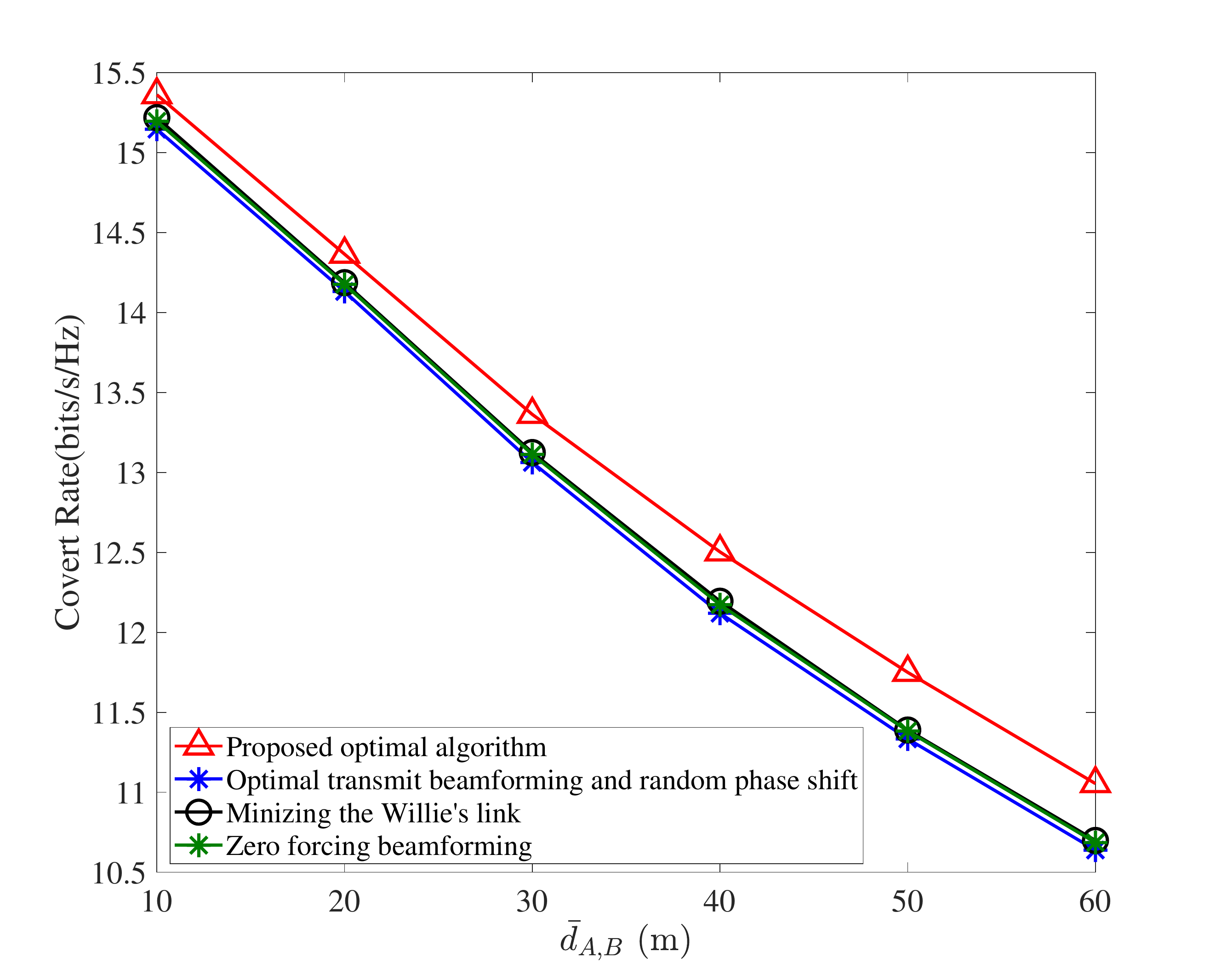}
\caption{The covert transmission rate against ${\bar{d}}_{A,B}$ with different algorithm, { $\rho=5$ $\mathbf{dB}$}, $\kappa=10^{-2}$, $M=5$, $N=20$,  $h_W=5$ m, $h_B=20$ m, $\mu_{A,B}=3$, $\mu_{A,S}=\mu_{S,B}=2$, $\mu_{A,W}=4$, $\mu_{S,W}=2$,  and ${\bar{d}}_{A,W}={d}_{A,S}=40$ m.} \label{Fig61}
\end{figure}

\begin{figure}
\centering
\includegraphics[width=0.4\textwidth]{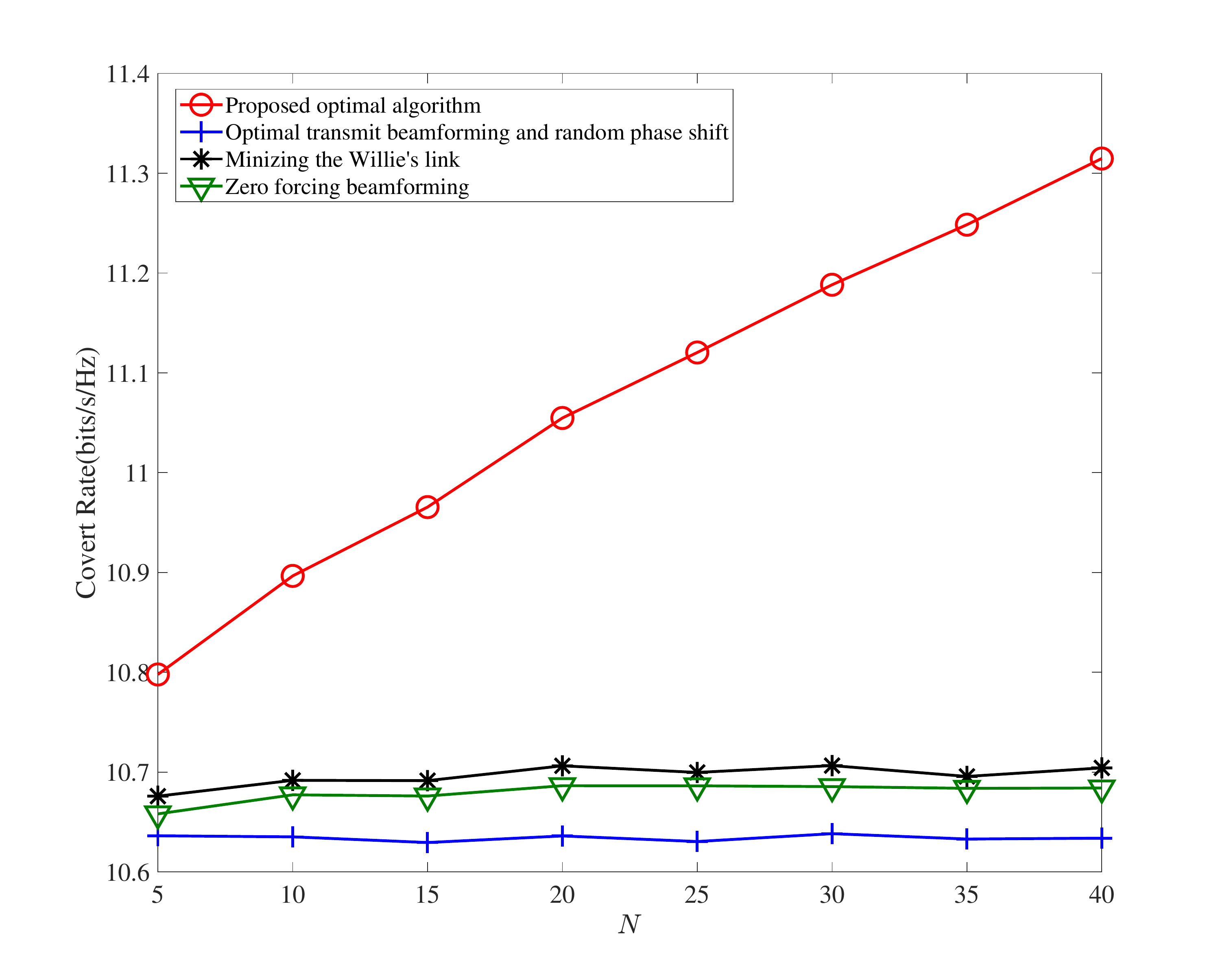}
\caption{The covert transmission rate against $N$ with instantaneous CSI of Willie's link, { $\rho=5$ $\mathbf{dB}$}, $\kappa=10^{-2}$, $M=5$,  $h_W=5$ m, $h_B=20$ m, $\mu_{A,B}=3$, $\mu_{A,S}=\mu_{S,B}=2$, $\mu_{A,W}=4$, $\mu_{S,W}=2$, ${\bar{d}}_{A,B}=60$ m, and ${\bar{d}}_{A,W}={d}_{A,S}=40$ m.}
\label{Fig71}
\end{figure}

In Fig. \ref{Fig61} and Fig. \ref{Fig71}, the covert transmission rates for the optimal and suboptimal algorithms proposed in this paper are plotted against $\bar{d}_{A,B}$ and against $N$. In terms of the covert transmission rate, the optimal algorithm is better than the algorithm minimizing Willie's link power gain, because the optimal algorithm achieves a trade off between maximizing  Bob's link and minimizing Willie's link power gains. The zero-forcing algorithm is the worse than the optimal algorithm and the suboptimal algorithm minimizing the Willie's link, because the diversity order of zero-forcing algorithm decreases to $M-1$, while the diversity order of the optimal algorithm and the algorithm minimizing Willie's link power gain is $M$. In addition, we can see that our proposed optimal algorithm is better than the optimal transmit beamforming and random phase shift algorithm.

\begin{figure}
\centering
\includegraphics[width=0.4\textwidth]{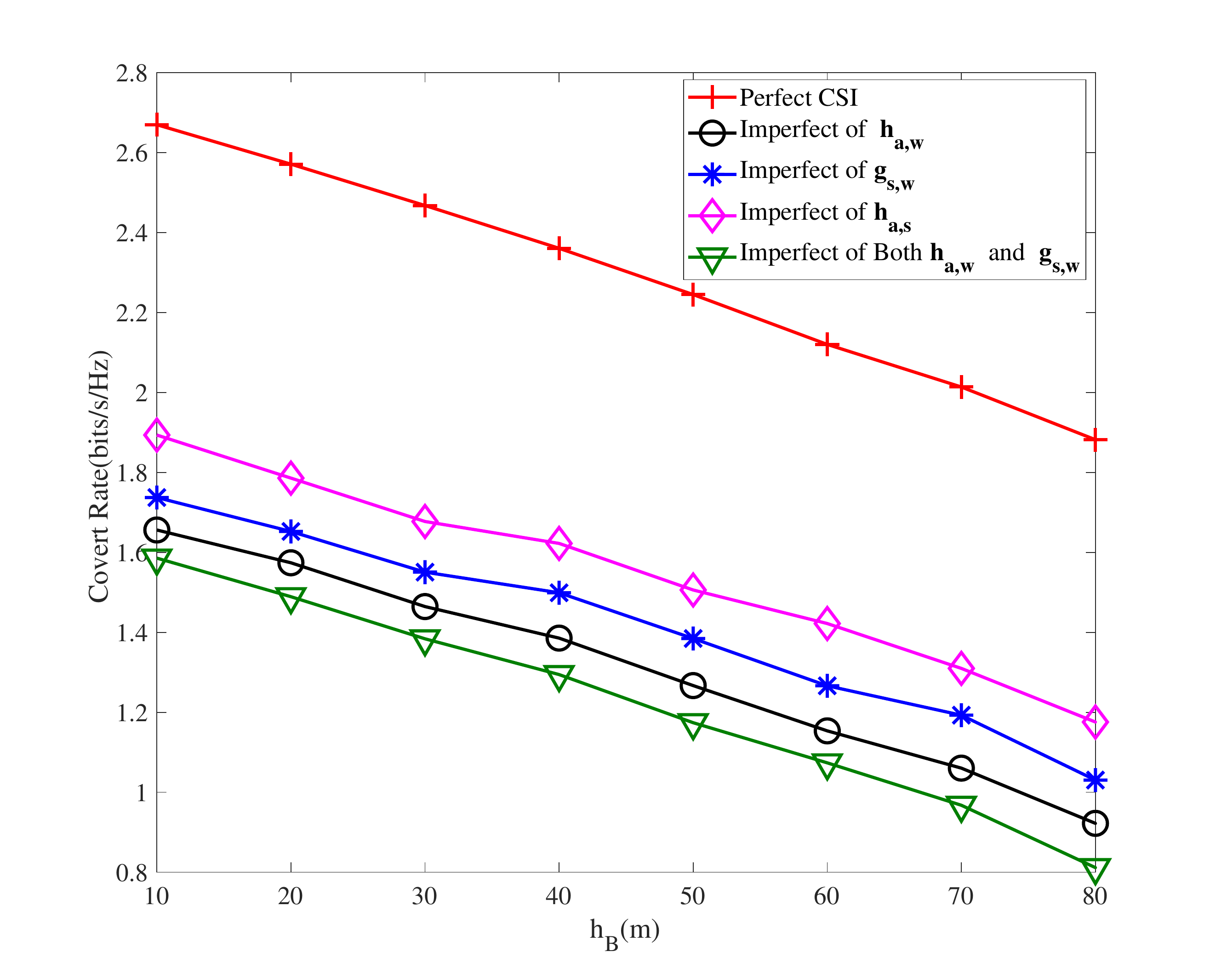}
\caption{The impact of imperfect CSI of the Willie's link on the covert transmission rate, ${\bar{d}}_{A,B}=60$ m, ${\bar{d}}_{S,B}=20$ m, $\rho=3$ $\mathbf{dB}$,, $\kappa=0.01$, $N=20$, $\mu_{A,B}=\mu_{S,B}=2$, $\mu_{A,S}=\mu_{A,W}=\mu_{S,W}=3$, ${\zeta}_{A,W}=5*{10}^{-9}$, ${\zeta}_{A,S}={\zeta}_{S,W}=5*{10}^{-6}$, $M=6$, $N=20$, and $d_{A,S}=d_{A,W}=d_{S,W}=40$ m.} \label{Fig8}
\end{figure}

The covert transmission rates with imperfect CSI of  Willie's link are shown in Fig. 8, where the distance between Alice, Willie, and the IRS is equal, i.e. $d_{A,S}=d_{A,W}=d_{S,W}=40$m. We find that the imperfect CSI of Willie's link leads to degradation of the covert transmission rate. In addition, we can see that in this scenario the covert rates decrease with the vertical distance $h_B$. {Moreover, the imperfect CSI of both Alice-to-Willie and IRS-to-Willie links has the most serious effect on the covert rate.  The imperfect CSI of Alice-to-Willie link has the second most serious effect on covert rate. This is because the Alice-to-Willie link is an independent part of Willie's link. By contrast, the combination of the Alice-to-IRS link and the IRS-to-Willie link is an independent part of Willie's link. In addition, since the Alice-to-IRS link affects both Willie's link and Bob's link power gains, the impact of the Alice-to-IRS link on the covert rate is lower than that of the IRS-to-Willie link.}
\section{Conclusion}
In this paper, we investigated the covert transmission rate in an IRS-aided system, where an energy detection method is applied at Willie. We found that the IRS does not always benefit the covert transmission. This conclusion is different from the multiuser system and secrecy transmission system where the IRS always improves performance. The reason is that in a covert transmission system, the signal power received at Willie should be less than a predetermined threshold, which limits the transmitter and IRS abilities to improve the quality of Bob's link. Hence, when we use IRS in covert transmission, the distances relationship between Bob, IRS, and Willie should be carefully considered. {Moveover, although in this paper the proposed algorithm is used to handle the case of a single warden, the proposed algorithm can also handle the case of multiple wardens.}
\begin{appendices}
\section{The PDF of $z$}
 The complex-valued variables ${h_{a,{s_i}}}$ and ${g_{{s_i},w}}$  can be expressed as ${h_{a,{s_i}}}{\rm{ = }}{A_i} + j{B_i}$, and
${g_{{s_i},w}} = {C_i} + j{D_i}$, where ${A_i}=Re\left({h_{a,{s_i}}}\right)$ and ${B_i}=Im\left({h_{a,{s_i}}}\right)$ follow the Gaussian distribution with mean zero and variance $\frac{{\sigma ^2}_{a,s}}{2}$, while  ${C_i}=Re\left({h_{{s_i},w}}\right)$ and ${D_i}=Im\left({h_{{s_i},w}}\right)$ follow the Gaussian distribution with mean zero and variance $\frac{{\sigma ^2}_{s,w}}{2}$. Then, we can obtain
\begin{align}\label{optimal2}\
{h_{a,{s_i}}}{g_{{s_i},w}} = {A_i}{C_i} - {B_i}{D_i} + j\left( {{B_i}{C_i} + {A_i}{D_i}} \right).
\end{align}
In \eqref{optimal2}, we have
${\mathbb{E}}\left( {{A_i}{C_i}} \right){\rm{ = 0}},{\mathbb{E}}\left( {{{\left( {{A_i}{C_i}} \right)}^{\rm{2}}}} \right){\rm{ = }}\frac{1}{4}{\sigma ^2}_{a,s}{\sigma ^2}_{s,w}$,
${\mathbb{E}}\left( {{B_i}{D_i}} \right){\rm{ = 0}},{\mathbb{E}}\left( {{{\left( {{B_i}{D_i}} \right)}^{\rm{2}}}} \right){\rm{ = }}\frac{1}{4}{\sigma ^2}_{a,s}{\sigma ^2}_{s,w}$,
${\mathbb{E}}\left( {{B_i}{C_i}} \right){\rm{ = 0}}, {\mathbb{E}}\left( {{{\left( {{B_i}{C_i}} \right)}^{\rm{2}}}} \right){\rm{ = }}\frac{1}{4}{\sigma ^2}_{a,s}{\sigma ^2}_{s,w}$, and
${\mathbb{E}}\left( {{A_i}{D_i}} \right){\rm{ = 0}}, {\mathbb{E}}\left( {{{\left( {{A_i}{D_i}} \right)}^{\rm{2}}}} \right){\rm{ = }}\frac{1}{4}{\sigma ^2}_{a,s}{\sigma ^2}_{s,w}$.
Thus, when $N\rightarrow\infty$, applying the central limit theorem, we can obtain that $\sum\limits_{i = 1}^N {{e^{j{\theta _i}}}{h_{a,{s_i}}}{g_{{s_i},w}}}$ follows the complex Gaussian distribution with zero mean and variance $N{\sigma ^2}_{a,s}{\sigma ^2}_{s,w}$. Since ${h_{a,w}}$ follows the complex Gaussian distribution with zero mean and variance ${\sigma ^2}_{a,w}$, ${h_{a,w}}+\sum\limits_{i = 1}^N {{e^{j{\theta _i}}}{h_{a,{s_i}}}{g_{{s_i},w}}}$ follows the complex Gaussian distribution with zero mean and variance  $N{\sigma ^2}_{a,s}{\sigma ^2}_{s,w}+{\sigma ^2}_{a,w}$.  Thus, the PDF of $z = {\left| {{h_{a,w}}{\rm{ + }}\sum\limits_{i = 1}^N {{e^{j{\theta _i}}}{h_{a,{s_i}}}{g_{{s_i},w}}} } \right|^2}$ is exponential distributed, and it is given by
\begin{align}\label{pdfz}\
{f_z}\left( z \right) = \frac{{\rm{1}}}{{\left( {{\sigma ^2}_{s,w}{\rm{ + }}N{\sigma ^2}_{a,s}{\sigma ^2}_{s,w}} \right)}}\exp \left( { - \frac{z}{{\left( {{\sigma ^2}_{s,w}{\rm{ + }}N{\sigma ^2}_{a,s}{\sigma ^2}_{s,w}} \right)}}} \right).
\end{align}
\end{appendices}
\bibliography{IRS1}

\begin{thebibliography}{10}
\providecommand{\url}[1]{#1}
\csname url@samestyle\endcsname
\providecommand{\newblock}{\relax}
\providecommand{\bibinfo}[2]{#2}
\providecommand{\BIBentrySTDinterwordspacing}{\spaceskip=0pt\relax}
\providecommand{\BIBentryALTinterwordstretchfactor}{4}
\providecommand{\BIBentryALTinterwordspacing}{\spaceskip=\fontdimen2\font plus
\BIBentryALTinterwordstretchfactor\fontdimen3\font minus
  \fontdimen4\font\relax}
\providecommand{\BIBforeignlanguage}[2]{{%
\expandafter\ifx\csname l@#1\endcsname\relax
\typeout{** WARNING: IEEEtran.bst: No hyphenation pattern has been}%
\typeout{** loaded for the language `#1'. Using the pattern for}%
\typeout{** the default language instead.}%
\else
\language=\csname l@#1\endcsname
\fi
#2}}
\providecommand{\BIBdecl}{\relax}
\BIBdecl

\bibitem{EBasar}
E.~{Basar}, M.~{Di Renzo}, J.~{De Rosny}, M.~{Debbah}, M.~{Alouini}, and
  R.~{Zhang}, ``Wireless communications through reconfigurable intelligent
  surfaces,'' \emph{IEEE Access}, vol.~7, pp. 116\,753--116\,773, 2019.

\bibitem{Qwu3}
Q.~{Wu} and R.~{Zhang}, ``Towards smart and reconfigurable environment:
  Intelligent reflecting surface aided wireless network,'' \emph{IEEE Commun.
  Mag.}, vol.~58, no.~1, pp. 106--112, Jan. 2020.

\bibitem{MAlouini}
J.~{Ye}, S.~{Guo}, and M.~{Alouini}, ``Joint reflecting and precoding designs
  for ser minimization in reconfigurable intelligent surfaces assisted {MIMO}
  systems,'' \emph{Accepted by IEEE Trans. Wireless Commun.}, 2020.

\bibitem{SJin}
Y.~{Han}, W.~{Tang}, S.~{Jin}, C.~{Wen}, and X.~{Ma}, ``Large intelligent
  surface-assisted wireless communication exploiting statistical {CSI},''
  \emph{IEEE Trans. Veh. Technol.}, vol.~68, no.~8, pp. 8238--8242, Aug. 2019.

\bibitem{DLi}
D.~{Li}, ``Ergodic capacity of intelligent reflecting surface-assisted
  communication systems with phase errors,'' \emph{Accepted by IEEE Commun.
  Lett.}, 2020.

\bibitem{CHuang}
C.~{Huang}, A.~{Zappone}, M.~{Debbah}, and C.~{Yuen}, ``Achievable rate
  maximization by passive intelligent mirrors,'' in \emph{2018 IEEE
  International Conference on Acoustics, Speech and Signal Processing
  (ICASSP)}, Apr. 2018, pp. 3714--3718.

\bibitem{MAlouini2}
Q.~{Nadeem}, A.~{Kammoun}, A.~{Chaaban}, M.~{Debbah}, and M.~{Alouini},
  ``Asymptotic {Max-Min} {SINR} analysis of reconfigurable intelligent surface
  assisted {MISO} systems,'' \emph{Accepted by IEEE Trans Wireless Commun.},
  2020.

\bibitem{Hhan}
H.~{Han}, J.~{Zhao}, Z.~{Xiong}, D.~{Niyato}, M.~D. {Renzo}, Q.~{Pham}, and
  W.~{Lu}, ``Intelligent reflecting surface aided power control for
  physical-layer broadcasting,'' \emph{https://arxiv.org/abs/1910.14383}, 2019.

\bibitem{Qwu1}
Q.~{Wu} and R.~{Zhang}, ``Intelligent reflecting surface enhanced wireless
  network via joint active and passive beamforming,'' \emph{IEEE Transactions
  on Wireless Commun.}, vol.~18, no.~11, pp. 5394--5409, Nov. 2019.

\bibitem{Qwu2}
------, ``Beamforming optimization for wireless network aided by intelligent
  reflecting surface with discrete phase shifts,'' \emph{IEEE Trans. Commun.},
  vol.~68, no.~3, pp. 1838--1851, Mar. 2020.

\bibitem{Hanzo}
C.~{Pan}, H.~{Ren}, K.~{Wang}, W.~{Xu}, M.~{Elkashlan}, A.~{Nallanathan}, and
  L.~{Hanzo}, ``Multicell {MIMO} communications relying on intelligent
  reflecting surfaces,'' \emph{Accepted by IEEE Trans Wireless Commun.}, 2020.

\bibitem{Gzhang}
M.~{Cui}, G.~{Zhang}, and R.~{Zhang}, ``Secure wireless communication via
  intelligent reflecting surface,'' \emph{IEEE Wireless Commun. Lett.}, vol.~8,
  no.~5, pp. 1410--1414, Oct. 2019.

\bibitem{JShi}
Z.~{Chu}, W.~{Hao}, P.~{Xiao}, and J.~{Shi}, ``Intelligent reflecting surface
  aided multi-antenna secure transmission,'' \emph{IEEE Wireless Commun.
  Lett.}, vol.~9, no.~1, pp. 108--112, Jan. 2020.

\bibitem{RSchober}
X.~{Yu}, D.~{Xu}, and R.~{Schober}, ``Enabling secure wireless communications
  via intelligent reflecting surfaces,'' \emph{IEEE Globalcom}, 2019.

\bibitem{YLiang}
J.~{Chen}, Y.~{Liang}, Y.~{Pei}, and H.~{Guo}, ``Intelligent reflecting
  surface: A programmable wireless environment for physical layer security,''
  \emph{IEEE Access}, vol.~7, pp. 82\,599--82\,612, 2019.

\bibitem{wxu}
H.~{Shen}, W.~{Xu}, S.~{Gong}, Z.~{He}, and C.~{Zhao}, ``Secrecy rate
  maximization for intelligent reflecting surface assisted multi-antenna
  communications,'' \emph{IEEE Commun. Lett.}, vol.~23, no.~9, pp. 1488--1492,
  Sep. 2019.

\bibitem{HWang}
L.~{Dong} and H.~{Wang}, ``Secure {MIMO} transmission via intelligent
  reflecting surface,'' \emph{IEEE Wireless Commun. Lett.}, pp. 1--1, 2020.

\bibitem{XGuan}
X.~{Guan}, Q.~{Wu}, and R.~{Zhang}, ``Intelligent reflecting surface assisted
  secrecy communication: Is artificial noise helpful or not?'' \emph{IEEE
  Wireless Communications Letters}, vol.~9, no.~6, pp. 778--782, 2020.

\bibitem{7355562}
B.~A. {Bash}, D.~{Goeckel}, D.~{Towsley}, and S.~{Guha}, ``Hiding information
  in noise: Fundamental limits of covert wireless communication,'' \emph{IEEE
  Commun. Mag.}, vol.~53, no.~12, pp. 26--31, Dec. 2015.

\bibitem{7084182}
S.~{Lee}, R.~J. {Baxley}, M.~A. {Weitnauer}, and B.~{Walkenhorst}, ``Achieving
  undetectable communication,'' \emph{IEEE J. Sel. Top. Signal Process.},
  vol.~9, no.~7, pp. 1195--1205, Oct. 2015.

\bibitem{8379465}
S.~{Yan}, B.~{He}, X.~{Zhou}, Y.~{Cong}, and A.~L. {Swindlehurst},
  ``Delay-intolerant covert communications with either fixed or random transmit
  power,'' \emph{IEEE Trans. Inf. Forensics Security}, vol.~14, no.~1, pp.
  129--140, Jan. 2019.

\bibitem{7964713}
T.~V. {Sobers}, B.~A. {Bash}, S.~{Guha}, D.~{Towsley}, and D.~{Goeckel},
  ``Covert communication in the presence of an uninformed jammer,'' \emph{IEEE
  Trans. Wireless Commun.}, vol.~16, no.~9, pp. 6193--6206, Sep. 2017.

\bibitem{Xliao}
X.~{Liao}, J.~{Si}, J.~{Shi}, Z.~{Li}, and H.~{Ding}, ``Generative adversarial
  network assisted power allocation for cooperative cognitive covert
  communication system,'' \emph{IEEE Communications Letters}, vol.~24, no.~7,
  pp. 1463--1467, July 2020.

\bibitem{HJiang}
X.~{Lu}, E.~{Hossain}, T.~{Shafique}, S.~{Feng}, H.~{Jiang}, and D.~{Niyato},
  ``Intelligent reflecting surface enabled covert communications in wireless
  networks,'' \emph{IEEE Network}, vol.~34, no.~5, pp. 148--155, 2020.

\bibitem{JS2}
Z.~{Cheng}, J.~{Si}, Z.~{Li}, D.~{Wang}, and N.~{Al-Dhahir}, ``Covert
  surveillance via proactive eavesdropping under channel uncertainty,''
  \emph{https://arxiv.org/abs/2002.06323}, 2020.

\bibitem{BHe}
B.~{He}, S.~{Yan}, X.~{Zhou}, and V.~K.~N. {Lau}, ``On covert communication
  with noise uncertainty,'' \emph{IEEE Commun. Lett.}, vol.~21, no.~4, pp.
  941--944, Apr. 2017.

\bibitem{HQ}
H.~Q. {Ta} and S.~W. {Kim}, ``Covert communication under channel uncertainty
  and noise uncertainty,'' in \emph{Proc. IEEE Int. Conf. Commun. (ICC)}, May
  2019.

\bibitem{ATaha}
A.~{Taha}, M.~{Alrabeiah}, and A.~{Alkhateeb}, ``Enabling large intelligent
  surfaces with compressive sensing and deep learning,'' in \emph{Available:
  https://arxiv.org/abs/1904.10136}, 2019.

\bibitem{DMishra}
D.~{Mishra} and H.~{Johansson}, ``Channel estimation and low-complexity
  beamforming design for passive intelligent surface assisted miso wireless
  energy transfer,'' in \emph{ICASSP 2019 - 2019 IEEE International Conference
  on Acoustics, Speech and Signal Processing (ICASSP)}, 2019, pp. 4659--4663.

\bibitem{ZQHe}
Z.~{He} and X.~{Yuan}, ``Cascaded channel estimation for large intelligent
  metasurface assisted massive mimo,'' \emph{IEEE Wireless Commun. Lett.},
  vol.~9, no.~2, pp. 210--214, 2020.

\bibitem{ZWang}
Z.~{Wang}, L.~{Liu}, and S.~{Cui}, ``Channel estimation for intelligent
  reflecting surface assisted multiuser communications: Framework, algorithms,
  and analysis,'' \emph{IEEE Trans. Wireless Commun.}, vol.~19, no.~10, pp.
  6607--6620, 2020.

\bibitem{XYuan}
H.~{Liu}, X.~{Yuan}, and Y.~J.~A. {Zhang}, ``Matrix-calibration-based cascaded
  channel estimation for reconfigurable intelligent surface assisted multiuser
  mimo,'' \emph{IEEE J. Sel. Areas Commun.}, vol.~38, no.~11, pp. 2621--2636,
  2020.

\bibitem{LTran}
L.~{Tran}, M.~F. {Hanif}, and M.~{Juntti}, ``A conic quadratic programming
  approach to physical layer multicasting for large-scale antenna arrays,''
  \emph{IEEE Signal Proce. Lett.}, vol.~21, no.~1, pp. 114--117, Jan. 2014.

\bibitem{ZhiQuanLuo}
N.~D. {Sidiropoulos}, T.~N. {Davidson}, and {Zhi-Quan Luo}, ``Transmit
  beamforming for physical-layer multicasting,'' \emph{IEEE Trans. Signal
  Process.}, vol.~54, no.~6, pp. 2239--2251, June 2006.

\bibitem{RZhang}
R.~{Zhang} and Y.~{Liang}, ``Exploiting multi-antennas for opportunistic
  spectrum sharing in cognitive radio networks,'' \emph{IEEE J. Sel. Top.
  Signal Process.}, vol.~2, no.~1, pp. 88--102, Feb. 2008.

\bibitem{YPei}
Y.~{Pei}, Y.~{Liang}, L.~{Zhang}, K.~C. {Teh}, and K.~H. {Li}, ``Secure
  communication over {MISO} cognitive radio channels,'' \emph{IEEE Trans.
  Wireless Commun.}, vol.~9, no.~4, pp. 1494--1502, Apr. 2010.

\bibitem{KTPhan}
K.~T. {Phan}, S.~A. {Vorobyov}, N.~D. {Sidiropoulos}, and C.~{Tellambura},
  ``Spectrum sharing in wireless networks via {QoS}-aware secondary multicast
  beamforming,'' \emph{IEEE Trans. Signal Process.}, vol.~57, no.~6, pp.
  2323--2335, June 2009.

\end{thebibliography}
\end{document}